\documentclass[prd,preprint,singlecolumn,superscriptaddress,preprintnumbers,amsmath,amssymb,tightenlines,longbibliography,nofootinbib]{revtex4-2}

\usepackage{epsfig}  
\usepackage{graphicx}
\usepackage{float}
\usepackage{color}
\usepackage[dvipsnames]{xcolor}
\usepackage{float}
\usepackage{amsfonts}
\usepackage{amsmath}

\usepackage{slashed}
\usepackage{mathrsfs}
\usepackage{siunitx}
\usepackage{multirow}
\usepackage{slashed}
\usepackage{soul}
\usepackage{braket}
\usepackage{dsfont}
\usepackage[colorlinks,citecolor=blue,urlcolor=blue,bookmarks=false,hypertexnames=true]{hyperref}
\usepackage{orcidlink}
\usepackage{booktabs}

\large



\begin{document}

\title{Leptonic CP Phase Determination from Fisher Information in T2K and NO$\nu$A}

\author{Neetu Raj Singh Chundawat\orcidlink{0000000300926260}}
\email{chundawat@ihep.ac.cn}
\affiliation{Institute of High Energy Physics, Chinese Academy of Sciences, Beijing 100049, China}
\affiliation{Kaiping Neutrino Research Center, Guangdong 529386, China}
\author{Luis A. Delgadillo\orcidlink{0000000325482299}}
\email{ldelgadillof@ihep.ac.cn}
\affiliation{Institute of High Energy Physics, Chinese Academy of Sciences, Beijing 100049, China}
\affiliation{Kaiping Neutrino Research Center, Guangdong 529386, China}

\author{Yu-Feng Li\orcidlink{0000000222205248}}
\email{liyufeng@ihep.ac.cn}
\affiliation{Institute of High Energy Physics, Chinese Academy of Sciences, Beijing 100049, China}
\affiliation{School of Physical Sciences, University of Chinese Academy of Sciences, Beijing 100049, China}

\begin{abstract}
\vspace{3em}
The precise determination of the leptonic CP phase (\(\delta_{\rm CP}\)) remains one of the central objectives of current and future long-baseline (LBL) neutrino oscillation experiments. Quantum estimation theory provides a natural framework to quantify the ultimate precision limits for estimating physical parameters encoded in quantum states. In this work, we employ the quantum Fisher information to investigate how much information about \(\delta_{\rm CP}\) is intrinsically encoded in neutrino states and how efficiently it is extracted in present LBL experiments such as T2K and NO$\nu$A. We first analyze the intrinsic quantum sensitivity of neutrino and antineutrino states and demonstrate how matter effects generate a neutrino mass-ordering dependent information structure. To compare the intrinsic information content of the quantum state with the information experimentally accessible through flavor measurements, we compute the event-level Fisher information from reconstructed event spectra using Poisson statistics including backgrounds and systematics. We find that both experiments extract only a small fraction of the total information available in the underlying quantum state. This extraction efficiency becomes particularly suppressed near maximally CP-violating regions, where the reconstructed event spectra exhibit reduced sensitivity to small variations in \(\delta_{\rm CP}\).  We also apply the same framework to the reactor mixing angle $\theta_{13}$, providing a benchmark for comparison with the estimation of $\delta_{\rm CP}$. Our analysis provides a complementary information-theoretic perspective on precise estimation of oscillation parameters in LBL neutrino experiments.
\end{abstract}

\maketitle
\newpage

\section{INTRODUCTION}

The discovery of neutrino oscillations has provided the first definitive evidence for physics beyond the Standard Model (SM)~\cite{ParticleDataGroup:2024cfk, JUNO:2025gmd}. The Pontecorvo-Maki-Nakagawa-Sakata (PMNS) mixing matrix, which relates flavor and mass eigenstates, is parameterized by three mixing angles ($\theta_{12}, \theta_{13}, \theta_{23}$) and a single Dirac CP-violating phase $\delta_{\mathrm{CP}}$~\cite{Pontecorvo:1957qd, Maki:1962mu}. Hence, the observation of a non-zero $\delta_{\mathrm{CP}}$ would constitute a violation of CP symmetry in the lepton sector, a potential ingredient in explaining the matter-antimatter asymmetry of the universe.

Over last decades, accelerator-based neutrino experiments have played a central role in the development of particle physics~\cite{Schwartz:1960hg, Lee:1960qv}. The current generation of long-baseline (LBL) experiments, Tokai-to-Kamioka (T2K) in Japan ($L = 295$~km)~\cite{T2K:2019bcf} and the NuMI Off-Axis $\nu_e$ Appearance (NO$\nu$A) in the USA ($L = 810$~km)~\cite{NOvA:2021nfi}, are at the forefront of research in neutrino physics. While both experiments employ high-intensity $\nu_\mu$ beams and measure $\nu_e$ appearance, their distinct baselines and detector technologies yield complementary sensitivities. For instance, T2K uses a lower-energy beam ($\sim 0.6$~GeV) and is sensitive to the first oscillation maximum, whereas NO$\nu$A provides a higher-energy beam ($\sim 2$~GeV) where matter effects are significant.

 Recently, a slight tension has emerged in the combined interpretation of their data. The T2K collaboration reports a preference for $\delta_{\mathrm{CP}} \sim 1.5\pi$~\cite{T2K:2023smv} under normal neutrino mass ordering , while NO$\nu$A data favor $\delta_{\mathrm{CP}} \sim 0.9\pi$~\cite{NOvA:2023iam}. The latest joint analysis from the two experiments, while resolving some degeneracies, still yields constrained but potentially conflicting intervals~\cite{T2K:2025wet}. Some explanations for this discrepancy range from statistical fluctuations to systematic uncertainties in neutrino flux or cross-section models, or even hints of physics beyond the Standard Model (BSM) such as neutrino non-standard interactions~\cite{Denton:2020uda, Chatterjee:2020kkm}. Thus, the current situation in the neutrino sector therefore motivates a more fundamental question: how much information about the oscillation parameters can be extracted from a given experiment?

Information has appeared as an increasingly important concept in physics, particularly following the pioneering work of Jaynes~\cite{Jaynes:1957zza}. The Cram\'er-Rao bound~\cite{Rao:1945} states that the achievable precision in estimating a physical parameter is determined by the information encoded in the measurement outcomes and improves with increasing statistics. This information about the parameter can be quantitatively characterized by the Fisher information~\cite{Fisher:1925}. Recently, the extension of classical Fisher information to its quantum counterpart, known as Quantum Fisher Information (QFI), has emerged within the framework of quantum estimation theory~\cite{Helstrom:1969,Paris:2007xql,Liu:2019kav,Lu:2010mv,Pezze:2014dwa}. The QFI naturally quantifies the sensitivity of a quantum state to changes in a parameter and determines the ultimate precision of its estimation via the Quantum Cram\'er-Rao bound (QCRB). In quantum metrology, it sets the fundamental limit on the achievable precision using quantum resources, enabling sensitivities beyond classical bounds. More recently, its scope has expanded to diverse areas of physics, including neutrino physics~\cite{Nogueira:2016hta,Ignoti:2025rxr,Frugiuele:2026yeq,Chundawat:2026jjd,Yadav:2026lsx,Farooq:2026eap}. Apart from quantum estimation theory, several studies have explored the role of other quantum information theoretic quantities in neutrino oscillations~\cite{Blasone:2007ou,Blasone:2009hb,Alok:2014vxa,Banerjee:2015dwa,Formaggio:2016vhr,Fu:2017gxd,Naikoo:2017pli,Naikoo:2018qss,Naikoo:2019kbi,Ming:2020,Blasone:2021fze,Jha:2025,Chattopadhyay:2023vxm,Alam:2026,Alok:2024une,Bouri:2024zgv,Alok:2025qqr,Alok:2024jsd,Banerjee:2026}.

In neutrino oscillations, the measurement process can be formulated within the most general framework of quantum measurements, namely positive operator-valued measures (POVMs)~\cite{Nielsen:2010}. Within this description, neutrino detection is effectively represented as a set of flavor measurements corresponding to a restricted class of operators acting on the neutrino state. The Fisher information associated with this POVM quantifies the amount of information about the underlying oscillation parameters that is accessible through physically realizable measurements. In particular, for standard neutrino oscillation experiments, this corresponds to the information that can be extracted from flavor measurements alone.
 
In neutrino detectors, the comparison between the neutrino state and the observables is not a simple projection onto flavor eigenstates, but also involves detector-specific effects such as energy binning, finite energy resolution, backgrounds and systematic uncertainties. The Fisher information corresponding to the measurements not only characterizes the sensitivity of the parameters under consideration but also highlights the limitations imposed by the detector itself, as compared to the case of perfect flavor measurements. In this work, we address the question of how much information is extracted in LBL experiments such as NO$\nu$A and T2K by quantifying both the intrinsic and experimentally accessible information content associated with physical measurements, and studying how this information is related to the precision achieved in these experiments. Our approach provides a complementary perspective by analyzing the experimentally extractable information associated with neutrino oscillation parameters, particularly the leptonic CP-violating phase, and comparing it with the underlying QFI encoded in the propagated neutrino state for the same parameter.

The structure of the paper is organized as follows. In Sec.~\ref{sec:fisher}, we introduce the Fisher Information formalism and discuss its relevant properties. In Sec.~\ref{sec:method}, we describe the methodology adopted in the present work. Sec.~\ref{sec:setup} is devoted to the experimental configurations and simulation setup for the T2K and NO$\nu$A experiments. In Sec.~\ref{sec:results}, we present the results related to the estimation of CP phase, and in Sec.~\ref{sec:theta13} we compare with the results obtained for reactor mixing angle. Sec.~\ref{sec:bkgplussys} provides a comprehensive treatment of backgrounds  and systematic uncertainties for T2K and  NO$\nu$A, and quantifies the associated degradation of the experimentally accessible Fisher information following their inclusion. Finally, we summarize our main findings in Sec.~\ref{sec:conclusion}.

\section{Fisher Information}
\label{sec:fisher}
\subsection{Quantum Fisher Information}

A central result in estimation theory is the Cramér-Rao bound, which states that the uncertainty of any unbiased estimator $\hat{\alpha}$ is bounded from below as
\begin{equation}
\Delta \hat\alpha
\ge
\frac{1}{\sqrt{N\,F(\alpha)}}\;,
\label{eq:CRB_classical}
\end{equation}
where $N$ denotes the number of independent measurements and
\begin{equation}
F(\alpha) = \sum_x \frac{1}{P(x|\alpha)} \left( \frac{\partial P(x|\alpha)}{\partial \alpha} \right)^2\;,
\label{eq: FI}
\end{equation}
is the Fisher information. Here, the sum extends to all possible measurement outcomes $x$. The factor $1/\sqrt{N}$ in Eq.~\eqref{eq:CRB_classical} reflects the statistical improvement obtained from repeated independent measurements.

The Fisher information obtained from any measurement is bounded above by the QFI by maximizing over all the possible POVM in quantum mechanics, leading to the QCRB. The QFI depends only on the quantum state that describes the system and is independent of the specific measurement performed. It is therefore interpreted as the maximum amount of information about the parameter $\alpha$ that can, in principle, be extracted from the quantum state, i.e.,
\begin{equation}
F_Q[\hat{\rho}(\alpha)] = \max_{\{\hat{E}_x\}} F(\alpha)\;,
\end{equation}
where $\hat{E}_x$ are the projectors that correspond to different POVMs. For a pure state $|\psi(\alpha)\rangle$, the QFI takes the simplified form
\begin{equation}
F_Q^{\mathrm{pure}}(\alpha)
=
4\left(
\langle \partial_\alpha\psi|\partial_\alpha\psi\rangle
-
\left|\langle \psi|\partial_\alpha\psi\rangle\right|^2
\right),
\label{eq:pure_QFI}
\end{equation}
where $|\partial_\alpha\psi\rangle \equiv
\frac{\partial}{\partial\alpha}|\psi\rangle$. 

\subsection{ Convexity and Additivity}

Both the QFI and the classical Fisher information satisfy several important mathematical properties. We list two of the most relevant properties that are used in our analysis below.

\textit{Convexity:} For a statistical mixture of states $\hat{\rho}_i(\alpha)$ with probabilities $p_i$, satisfying $\sum_i p_i = 1$, the QFI obeys
\begin{equation}
F_Q\!\left(\sum_i p_i \hat{\rho}_i(\alpha)\right)
\le
\sum_i p_i \, F_Q\!\left(\hat{\rho}_i(\alpha)\right).
\label{eq:QFI_convexity_general}
\end{equation}
Similarly, the statistical mixture of states with weights $p_i$, the Fisher information satisfies
\begin{equation}
F(\alpha) \le \sum_i p_i \, F_i(\alpha).
\end{equation}
This reflects the fact that statistical mixing of quantum states does not enhance the attainable estimation precision.

\textit{Additivity:} For $N$ independent subsystems undergoing independent measurements or $N$ uncorrelated events, the total Fisher information is given by the sum of the individual contributions,
\begin{equation}
F(\alpha) = \sum_{i=1}^{N} F_i(\alpha),
\end{equation}
where $F_i(\alpha)$ represents the Fisher information associated with the $i$-th subsystem.

For identically prepared and independent systems, this expression simplifies to
\begin{equation}
F(\alpha) = N\,F^{(1)}(\alpha),
\end{equation}
where $F^{(1)}(\alpha)$ denotes the Fisher information for a single subsystem. As a result, the Cramér-Rao bound scales similar to as given in Eq.~\eqref{eq:CRB_classical}.

A similar additivity property holds for the QFI. For $N$ independent quantum states $\hat{\rho}^{(i)}(\alpha)$, one obtains
\begin{equation}
F_Q\!\left[\bigotimes_{i=1}^{N} \hat{\rho}^{(i)}(\alpha)\right]
=
\sum_{i=1}^{N} F_Q\!\left[\hat{\rho}^{(i)}(\alpha)\right].
\end{equation}

In the special case of $N$ identical copies $\hat{\rho}(\alpha)$, the total QFI reduces to
\begin{equation}
F_Q^{\mathrm{tot}}(\alpha) = N\,F_Q[\hat{\rho}(\alpha)].
\label{eq:additivity}
\end{equation}


\subsection{Flavor measurement as a POVM}

The neutrinos in the experiments are detected by the final flavour of charged lepton produced via charged current interactions of neutrinos. This measurement process can be described in quantum mechanics by a specific POVM, $\{\hat{E}_x\}$, associated with flavor detection. For a system described by $\hat{\rho}(\alpha)$, the probability to obtain $x$ can be written as, $P(x|\alpha) = \mathrm{Tr}\!\left[\hat{\rho}(\alpha)\hat{E}_x\right]$. Thus, all measurement outcomes and their associated probabilities are determined by the pair $\{\hat{\rho}(\alpha), \hat{E}_x\}$.

In many physics scenarios, any quantum measurement can be approximated as an effective two-outcome POVM ($\hat{E}_1, \, \hat{E}_2 = \mathbb{I} - \hat{E}_1,$), for this case, the Fisher information corresponding to the flavor measurement can be written using Eq.~\eqref{eq: FI} as, 
\begin{equation}
F_{l}(\alpha)
=
\frac{
\left(\partial_\alpha P_1(\alpha)\right)^2
}{
P_1(\alpha)\left[1-P_1(\alpha)\right]
},
\end{equation}
where $P_1 (\alpha)$ is probability corresponding to any projector $\hat{E}_1$. However, for more general probability distributions involving multiple outcomes, the corresponding Fisher information must be computed using the full expression given in Eq.~\eqref{eq: FI}. The QFI is defined as the maximum of Fisher information calculated over all possible POVMs and satisfies, 
\begin{equation}
F_{l}(\alpha) \le F_Q^{}(\alpha),
\end{equation}

where the equality holds for optimal measurement strategies. Neutrino detection relies primarily on flavor identification via charged-current weak interactions, which effectively restricts the accessible measurements to the flavor basis. In contrast, for other quantum systems, more general measurement strategies can, in principle, be implemented to extract a larger amount of information from the underlying quantum state. It is also clear from above equation that flavor POVM corresponds to an experimentally realizable measurement, while the QFI is an ideal upper bound derived from neutrino state.

\section{Methodology}
\label{sec:method}
The LBL neutrino oscillation experiments NO$\nu$A and T2K play a crucial role not only in addressing open questions within the standard three-neutrino framework but also in probing a variety of BSM physics scenarios~\cite{Machado:2011jt, T2K:2019jwa, Chen:2015siy, Agarwalla:2016mrc, Forero:2016cmb, Miranda:2016wdr, Pasquini:2016kwk, Ghosh:2017atj, Rout:2017udo, Brdar:2017kbt, Capozzi:2019iqn, deGouvea:2022kma, Ngoc:2022uhg, Alok:2022jxo, Cherchiglia:2023ojf, Li:2023mil, Konwar:2024nwc, NOvA:2024lti, Yu:2024nkc, Ternes:2024qui, Alonso-Alvarez:2024wnh, Alves:2026ydc}. These experiments operate at characteristic $L/E$ values where oscillations driven by the atmospheric mass-squared difference dominate and accurate description of neutrino propagation requires matter effects taken into consideration. The evolution of an initial three-flavor state neutrino is governed by the effective Hamiltonian in matter,

\begin{equation}
H = \frac{1}{2E} U \,\mathrm{diag}(0,\Delta m_{21}^2,\Delta m_{31}^2)\,U^\dagger + \mathrm{diag}(V_{CC},0,0),
\end{equation}
where $U$ is the PMNS matrix and $V_{CC}$ is the matter potential. The matter potential is computed as:
\begin{equation}
V_{CC} = \sqrt{2} G_F n_e = 7.56 \times 10^{-14}~\text{eV} \cdot \rho\,[\text{g/cm}^3] \cdot Y_e,
\end{equation}
where $G_F$ is the Fermi constant and $n_e$ is the electron number density.

The neutrino state evolution is obtained by $|\nu(L)\rangle = e^{-iHL}|\nu(0)\rangle$, where $L$ is the propagation length. The appearance probability of $\nu_\mu \to \nu_e$ can be given by $P_{\mu e} = \left| \langle \nu_e | \nu(L)\rangle \right|^2$. For antineutrinos, within the particle data group parameterization (PDG) of the leptonic mixing matrix~\cite{ParticleDataGroup:2024cfk}, the CP phase transforms as $\delta_{\rm CP}\to -\delta_{\rm CP}$ and the sign of matter potential will be flipped~\cite{Denton:2024thm}. In our analysis, we consider the oscillation probability expressions in the approximation for constant matter density with $\rho_{\text{NO$\nu$A}} = 2.8$~g/cm$^3$, $\rho_{\text{T2K}} = 2.6$~g/cm$^3$, and electron fraction $Y_e = 0.5$. The primary physics goals of current and future LBL experiments include the precise determination of $\delta_{\rm CP}$ and the resolution of the neutrino mass ordering~\cite{Huber:2004gg, Nunokawa:2005nx, Li:2013zyd, JUNO:2015zny, DUNE:2020ypp, Hyper-Kamiokande:2018ofw, Alekou:2022emd, Blennow:2013swa, Cabrera:2020ksc, Delgadillo:2023lyp, Parke:2024xre, Goswami:2025wla}, which is directly linked to the sign of the atmospheric mass squared splitting $\Delta m^2_{31}$.

Recently, studies based on quantum estimation theory have investigated the QFI associated with $\delta_{\rm CP}$ also within the single-parameter estimation framework and compared it with the Fisher information obtained from flavor POVMs, highlighting the role of flavor measurements in limiting the accessible CP information~\cite{Ignoti:2025rxr, Yadav:2026lsx, Frugiuele:2026yeq}. The information contained in the $\nu_\mu \rightarrow \nu_e$ appearance channel can, in principle, be quantified using the appearance probability $P_{\mu e}(E_k,\delta_{\rm CP})$. However, in the present analysis, we do not separately evaluate the Fisher information associated with the flavour POVM, since long-baseline experiments do not directly measure the oscillation probability at a given energy. Instead, these oscillation probabilities are embedded in the reconstructed experimental event spectra, which include neutrino fluxes, interaction cross sections, and detector effects such as energy smearing, detector efficiencies, and backgrounds. Hence, the comparison between the event-level Fisher information and the event-weighted QFI benchmark provides a more reliable framework for understanding the experimentally accessible quantum information.

For a single neutrino, the QFI is given by Eq.~\eqref{eq:pure_QFI}. Let us now consider a LBL neutrino experiment with an expected number of events $N_k(\delta_{\rm CP})$ in the $k$-th energy bin. To establish the quantum-information benchmark, we estimate the number of neutrinos in the propagated ensemble prior to flavor conversion. The unoscillated event spectra in each energy bin is computed as
\begin{equation}
    C_k=\frac{N_k(\delta_{\rm CP})}{P_{\mu e}(E_k,\delta_{\rm CP})}\;.
    \label{eq:unosc}
\end{equation}
We have verified that the difference between this bin-center approximation and the corresponding flux--cross-section convolution is negligible for the adopted energy binning and detector configurations.
 Using the additivity property of QFI, the corresponding QFI contribution per energy bin relevant for the detector can be approximated using Eqs.~\eqref{eq:pure_QFI},~\eqref{eq:additivity} and \eqref{eq:unosc} as

\begin{equation}
\label{eq:wqfi}
F_{Q}^{(k)}=C_k F_Q^{\rm pure}(E_k,\delta_{\rm CP})\;.
\end{equation}
The total event-weighted QFI is then obtained by summing over all reconstructed energy bins,
\begin{equation}
\label{eq:totqfi}
F_Q^{\rm tot}(\delta_{\rm CP})
=
\sum_k
F^{(k)}_{Q}(\delta_{\rm CP})\;.
\end{equation}

Now to calculate the Fisher information at the event-level, we use standard statistics formalism~\cite{Cowan:1998ji}. Assuming statistically independent Poisson distributed events in each bin, the Fisher information associated with the parameter $\delta_{\rm CP}$ is given by,

\begin{equation}
F_{\rm event}(\delta_{\rm CP})
=
\left\langle
\left(
\frac{\partial \ln \mathcal{L}}
{\partial \delta_{\rm CP}}
\right)^2
\right\rangle\;,
\label{eq:log}
\end{equation}
where $\mathcal{L}$ is the log-likelihood. Since the standard $\chi^2$ function is $\chi^2(\delta_{\rm CP}) = -2\ln \mathcal{L}$ then the Fisher information corresponds to the local curvature of the $\chi^2$ function around the true values of parameter into consideration.

Using the general properties of Poisson statistics in Eq.~\eqref{eq:log}, the event-level Fisher information per energy bin $k$, can be written as, 

\begin{equation}
F^{(k)}_{\rm event}(\delta_{\rm CP})
=
\frac{
\left(
\partial_{\delta_{\rm CP}}
N_k(\delta_{\rm CP})
\right)^2
}{
N_k(\delta_{\rm CP})
}\;,
\label{eq:Fevent}
\end{equation}

and the total event-level Fisher information can be given by summing over all the energy bins,

\begin{equation}
F^{\rm tot}_{\rm event}(\delta_{\rm CP})
=
\sum_k
F^{(k)}_{\rm event}(\delta_{\rm CP})\;.
\end{equation}
Having established the event-level Fisher information formalism, in the next section, we now describe the experimental configurations relevant for this study, namely for NO$\nu$A and T2K experiments.

\section{Experimental Setup}
\label{sec:setup}
In this section of the manuscript, we will focus on the description of the experimental configurations relevant for this study. Our analysis is specifically focused on the NO$\nu$A and T2K experimental configurations, these LBL experiments operate at different baselines and energy regimes, thereby offering complementary sensitivities to oscillation parameters and potential new physics searches. In our simulations, we employ the \textsc{GLoBES} (General Long Baseline Experiment Simulator) software package~\cite{Huber:2004ka, Huber:2007ji}, which provides a  comprehensive framework for simulating neutrino oscillation experiments including detector responses, neutrino fluxes, and systematic uncertainties. For each experiment, we employ the standard configuration files distributed within~\textsc{GLoBES}, which were adapted to reproduce the nominal experimental setups as described in Refs.~\cite{T2K:2019bcf, NOvA:2021nfi}. These configurations contain experiment-specific parameters such as baseline lengths, target masses, neutrino fluxes, cross-sections, energy resolutions, and efficiencies. In the following subsections we detail the specific input parameters adopted for the NO$\nu$A and T2K simulations, respectively.

\subsection{T2K Experimental Configuration}

The T2K experimental configuration provides a well-established framework for LBL neutrino oscillation studies, it employs a 295~km baseline from the J-PARC facility. In our analysis of the T2K setup we consider a fiducial target mass of 25~kt with a total exposure of 10 years, 4 years in neutrino mode and 6 years in antineutrino mode, driven by a beam power of 0.77 MW. Besides, the corresponding reconstructed events are selected within an energy range of 0.4 GeV to 1.2 GeV, binned into 20 equally distributed intervals. Furthermore, an energy resolution of 14\% is assumed for both muon and electron channels. Here, the corresponding neutrino and antineutrino fluxes are normalized to $6.93$ with a stored muon rate of $10^{20}$ protons-on-target (POT) per year. In Table~\ref{tab:t2k_config} we summarize the relevant parameters assumed in our simulation. In addition, the cross sections for charged-current (CC), neutral-current (NC), and charged-current quasi-elastic (CCQE) interactions are adopted from the corresponding GLoBES implementation. For the appearance channel, the CCQE sample is used to extract spectral information, whereas the CC sample provides rate information. The corresponding channel configurations are summarized in Table~\ref{tab:t2k_channels}.

\begin{table}[H]
\centering
\caption{T2K experimental configuration parameters.}
\label{tab:t2k_config}
\begin{tabular}{llc}
\toprule
\textbf{Parameter} & \textbf{Description} & \textbf{Value} \\
\hline \hline
$L$ & Baseline length & 295.0 km \\
$M_{\text{target}}$ & Fiducial mass & 25.0 kt \\
$t_{\nu}$ & Neutrino running time & 4 years \\
$t_{\bar{\nu}}$ & Antineutrino running time & 6 years \\
$P_{\text{beam}}$ & Beam power & 0.77 MW \\
$E_{\text{min}}$ & Minimum reconstructed energy & 0.4 GeV \\
$E_{\text{max}}$ & Maximum reconstructed energy & 1.2 GeV \\
$N_{\text{bins}}$ & Number of energy bins & 20 \\
$\Delta E_\nu/E_\nu$ & Energy resolution ($\mu$ channel) & 14\% \\
$\Delta E_\nu/E_\nu$ & Energy resolution ($e$ channel) & 14\% \\
\hline
\end{tabular}
\end{table}

As far as background contaminants are concerned, the main background contributions arise from three sources. The first source of background is due to intrinsic $\nu_e$ beam contamination, which is approximately $0.5\%$ of the expected $\nu_e$-CC. The next source of contaminants consists of NC events misidentified as $\nu_e$ interactions, contributing about $0.56\%$. The last source of background arises from $\nu_\mu$-CC events accompanied by $\pi^0$ production, which represent a minor fraction of about $0.03\%$. We also account for wrong-sign (WS) beam contamination in the appearance channel for antineutrino-mode running, while neglecting the corresponding contribution in neutrino mode. The oscillated WS neutrinos serve as an important background contribution. The T2K Collaboration predicts approximately $3.0$ WS $\nu_\mu\rightarrow\nu_e$ events in the antineutrino-mode sample~\cite{T2K:2019ird}, which we use for the normalization in our simulation. The corresponding WS contribution in neutrino-mode running is considerably smaller and is therefore neglected in the present analysis for the T2K configuration.

In addition to the background contributions, we incorporate systematic
uncertainties associated with the signal and background normalizations, parameterized following Ref.~\cite{Huber:2002mx}. The normalization uncertainties adopted for the different signal channels, together with their corresponding efficiencies, are summarized in Table~\ref{tab:t2k_channels}, while a $20\%$ normalization uncertainty is assigned to the aforementioned background components. In addition, we include a $5\%$ energy-calibration uncertainty following
Ref.~\cite{Goswami:2025wla}.

\begin{table}[H]
\centering
\caption{T2K signal channels and systematic uncertainties.}
\label{tab:t2k_channels}
\begin{tabular}{lcc}
\toprule
\textbf{Channel} & \textbf{Signal Efficiency} & \textbf{Normalization Error} \\
\hline \hline
$\nu_\mu \to \nu_e$ (CCQE) & 50.5\% & 10\% \\
$\bar{\nu}_\mu \to \bar{\nu}_e$ (CCQE) & 50.5\% & 10\% \\
$\nu_\mu \to \nu_\mu$ (CCQE) & 90\% & 2.5\% \\
$\bar{\nu}_\mu \to \bar{\nu}_\mu$ (CCQE) & 90\% & 2.5\% \\
$\nu_\mu \to \nu_e$ (CC) & 50.5\% & 5\% \\
$\bar{\nu}_\mu \to \bar{\nu}_e$ (CC) & 50.5\% & 5\% \\
\hline
\end{tabular}
\end{table}

\subsection{NO$\nu$A Experimental Configuration}
The NO$\nu$A experiment features a longer baseline of 812~km from Fermilab to a 25~kt detector in Ash River, Minnesota. In this analysis, we consider a fiducial target mass of 25~kt with a total exposure of 10 years, 5 years in neutrino mode and 5 years in antineutrino mode, driven by a beam power of 1.12~MW. Besides, the reconstructed events are selected within an energy range of 1.0~GeV to 3.5~GeV, binned into 20 equally-distributed intervals. An energy resolution of 16\% is assumed for the muon channel and 14\% for the electron channel, reflecting the different reconstruction performances for each topology. The fluxes for this configuration are properly normalized to $6.48 \times 10^{-19}$ with a stored muon rate of $10^{20}$ POT per-year. In Table~\ref{tab:nova_config} we display the corresponding parameters considered in our simulation of the NO$\nu$A configuration.

\begin{table}[H]
\centering
\caption{NO$\nu$A experimental configuration parameters.}
\label{tab:nova_config}
\begin{tabular}{llc}
\toprule
\textbf{Parameter} & \textbf{Description} & \textbf{Value} \\
\hline \hline
$L$ & Baseline length & 812.0 km \\
$M_{\text{target}}$ & Fiducial mass & 25.0 kt \\
$t_{\nu}, t_{\bar{\nu}}$ & Running time (each) & 5 years \\
$P_{\text{beam}}$ & Beam power & 1.12 MW \\
$E_{\text{min}}$ & Minimum reconstructed energy & 1.0 GeV \\
$E_{\text{max}}$ & Maximum reconstructed energy & 3.5 GeV \\
$N_{\text{bins}}$ & Number of energy bins & 20 \\
$\Delta E_\nu/E_\nu$ & Energy resolution ($\mu$ channel) & 16\% \\
$\Delta E_\nu/E_\nu$ & Energy resolution ($e$ channel) & 14\% \\
\hline
\end{tabular}
\end{table}

The background budget comprises three main contributions, the largest of which arises from the intrinsic $\nu_e$ beam contamination, which accounts for about $12.0\%$ of the $\nu_e$-CC events. Moreover, misidentified $\nu_\mu$-NC events with $\pi^0$ production contribute $0.15\%$, while the charged-current $\nu_\mu$ misidentification events represent a smaller fraction of about $0.04\%$. For the NO$\nu$A configuration, we also account for WS beam contamination in the appearance channel for both beam modes, with the corresponding appearance contributions normalized to the reported expectations of $1.0$ and $2.3$ events in neutrino and antineutrino modes, respectively~\cite{NOvA:2021nfi}. The corresponding normalization uncertainties adopted for the different signal channels, together with their corresponding efficiencies, are summarized in Table~\ref{tab:nova_channels}, for the background components, we assigned a $20\%$ normalization uncertainty. Besides, we include a $5\%$ energy-calibration uncertainty as in the T2K configuration. In Table~\ref{tab:nova_channels} we present the corresponding efficiencies and normalization uncertainties for the expected signal channels adopted in our simulation of the NO$\nu$A setup~\cite{Yu:2024nkc, NOvA:2025tmb}.

\begin{table}[H]
\centering
\caption{NO$\nu$A signal channels and systematic uncertainties.}
\label{tab:nova_channels}
\begin{tabular}{lcc}
\hline \hline
\textbf{Channel} & \textbf{Signal Efficiency} & \textbf{Normalization Error}  \\
\hline \hline
$\nu_\mu \to \nu_e$ (CC) & 24\% & 5\%  \\
$\nu_\mu \to \nu_\mu$ (CC) & 80\% & 5\%  \\
$\bar{\nu}_\mu \to \bar{\nu}_e$ (CC) & 37\% & 5\%  \\
$\bar{\nu}_\mu \to \bar{\nu}_\mu$ (CC) & 80\% & 5\%  \\
\hline
\end{tabular}
\end{table}

We first compute the expected signal event spectrum for each beam polarity. The background contributions and systematic effects, including normalization and energy-calibration uncertainties, are described in Sec.~\ref{sec:bkgplussys}.

\subsection{Expected event rates computation in T2K and NO$\nu$A}
The oscillation probabilities are computed including matter effects under the constant density approximation. In our simulations, we adopt the best-fit values of the neutrino oscillation parameters from Ref.~\cite{Esteban:2024eli}, which are summarized in Table~\ref{tab:para}. In the following discussions, NO and IO denote normal ordering and inverted ordering, respectively. Besides, the expected number of events in energy bin $k$ for a given channel is computed as:
\begin{equation}
N_k(\delta_{\mathrm{CP}}) = \int_{E_\nu^{\text{min}}}^{E_\nu^{\text{max}}} \Phi(E_\nu) \, \sigma(E_\nu) \, P_{\mu e}(E_\nu, \delta_{\mathrm{CP}}) \, \epsilon(E_\nu) \, R(E_\nu, E_k) \, dE_\nu,
\label{eq:event_rate}
\end{equation}
where $\Phi(E_\nu)$ is the neutrino flux, $\sigma(E_\nu)$ is the interaction cross section, $P_{\mu e}(E_\nu, \delta_{\mathrm{CP}})$ is the oscillation probability including matter effects, $\epsilon(E_\nu)$ is the detection efficiency, and $R(E_\nu, E_k)$ is the energy resolution function, modeled as Gaussian smearing with width $\sigma$ taken from Table~\ref{tab:t2k_config} for T2K or Table~\ref{tab:nova_config} for NO$\nu$A.

\begin{table}[H]
\centering
\caption{Corresponding neutrino oscillation parameters considered in this study.}
\begin{tabular}{l c}
\toprule
\label{tab:para}
Neutrino Oscillation Parameter & Best-fit value \\
\hline \hline
$\theta_{12}$ & $33.44^\circ$ \\
$\theta_{13}$ & $8.57^\circ$ \\
$\theta_{23}$ & $49.2^\circ$ \\
$\Delta m_{21}^2$ & $7.42 \times 10^{-5}~\text{eV}^2$ \\
$\Delta m_{31}^2~(\text{NO})$ & $+2.517 \times 10^{-3}~\text{eV}^2$ \\
$\Delta m_{31}^2~(\text{IO})$ & $-2.498 \times 10^{-3}~\text{eV}^2$ \\
\hline 
\end{tabular}
\end{table}

In Table~\ref{tab:validation}, we summarize the total (simulated) expected event rates against the published results from the T2K and NO$\nu$A collaborations, including both appearance and disappearance channels. These rates correspond to the combined contributions from neutrino and antineutrino polarities. For each experimental configuration and each value of $\delta_{\mathrm{CP}}$ displayed, the comparison between our simulation and the measured events leads to differences below $5\%$.

\begin{table}[H]
\centering
\caption{Total (appearance plus disappearance) expected event rates.}
\label{tab:validation}
\begin{tabular}{lcccc}
\toprule
\textbf{Experiment} & $\delta_{\mathrm{CP}}$ & \textbf{Our Simulation} & \textbf{Reference} & \textbf{Difference} \\
\hline \hline
T2K (NO) & $0$ & $78.4 \pm 8.9$ & $75.3 \pm 9.2$ \cite{T2K:2019bcf} & 4.1\% \\
T2K (NO) & $\pi$ & $124.2 \pm 11.1$ & $118.7 \pm 10.8$ \cite{T2K:2019bcf} & 4.6\% \\
NO$\nu$A (NO) & $0$ & $42.3 \pm 5.1$ & $40.8 \pm 5.5$ \cite{NOvA:2021nfi} & 3.7\% \\
NO$\nu$A (NO) & $\pi$ & $28.7 \pm 4.2$ & $27.9 \pm 4.5$ \cite{NOvA:2021nfi} & 2.9\% \\
\hline 
\end{tabular}
\end{table}

\begin{figure}[htb!]
    \centering
    \includegraphics[width=1\linewidth]{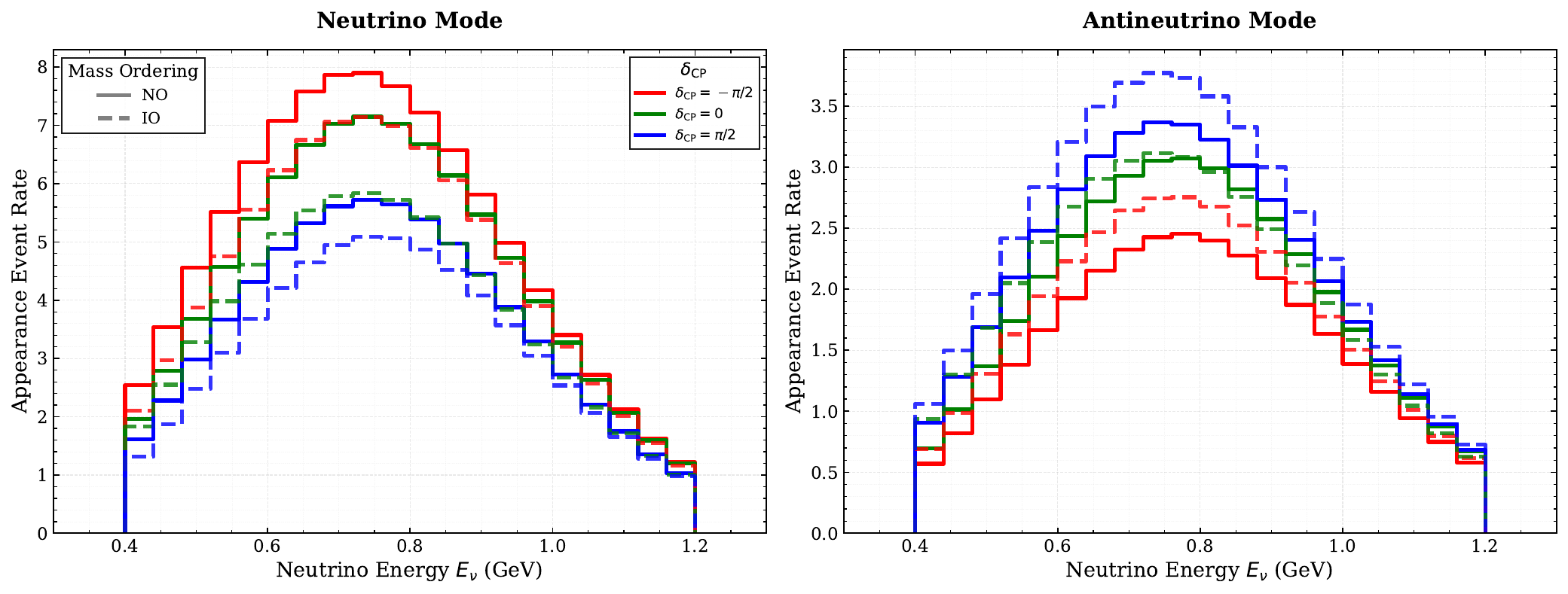}
    \caption{Expected $\nu_e$ (solid lines) and $\bar{\nu}_e$ (dashed lines) appearance event spectra for the T2K configuration, evaluated at three representative values of $\delta_{\mathrm{CP}}$ ($0$, $-\pi/2$, and $\pi/2$). The spectra correspond to a total exposure of 4 years in neutrino mode and 6 years in antineutrino mode with a 0.77~MW beam, accumulated within the reconstructed energy range of 0.4 to 1.2~GeV. See the text for a detailed explanation.}
    \label{fig:t2k_spectrum}
\end{figure}

In Fig.~\ref{fig:t2k_spectrum}, we present the expected neutrino  and antineutrino  appearance event spectra for the T2K experimental configuration. The event rates are evaluated at $\delta_{\mathrm{CP}} = 0$, $-\pi/2$, and $\pi/2$, providing a comprehensive comparison between CP-conserving and CP-violating phase values. For both neutrino and antineutrino modes, the event spectrum peaks near 0.7~GeV, consistent with the flux maximum at the 295~km baseline in T2K. The spectral separation is notable between the $\delta_{\mathrm{CP}} = 0$ (green lines) and $\delta_{\mathrm{CP}} = \pm\pi/2$ (red/blue lines) values, in both neutrino and antineutrino channels, illustrating the sensitivity of the T2K experiment to leptonic CP violation.
\begin{figure}[htb!]
    \centering
    \includegraphics[width=1\linewidth]{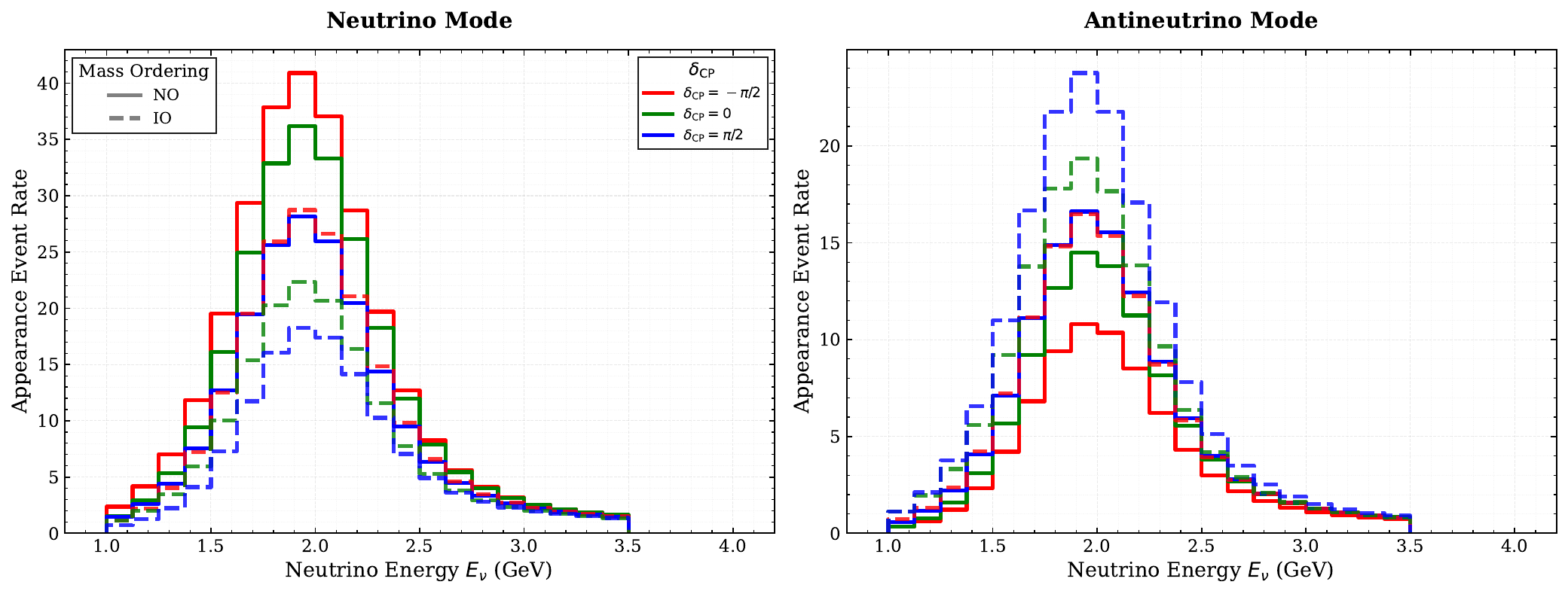}
    \caption{Expected $\nu_e$ (solid lines) and $\bar{\nu}_e$ (dashed lines) appearance event spectra for the NO$\nu$A configuration, evaluated at three representative values of $\delta_{\mathrm{CP}}$ ($0$, $-\pi/2$, and $\pi/2$). The spectra correspond to a total exposure of 5 years in neutrino mode and 5 years in antineutrino mode with a 1.12~MW beam, accumulated within the energy window of 1.0 to 3.5~GeV. See the text for a detailed explanation.}
    \label{fig:nova_spectrum}
\end{figure}

In Fig.~\ref{fig:nova_spectrum}, we display the expected neutrino and antineutrino appearance events for the NO$\nu$A configuration. The event distributions are evaluated at three benchmark values of the CP-violating phase: $\delta_{\mathrm{CP}} = 0$, $-\pi/2$, and $\pi/2$, including both CP-conserving and CP-violating values. The corresponding spectral peak around 2~GeV is consistent with the oscillation maximum for the 812~km baseline at NO$\nu$A. For instance, a distinct separation is observed between the $\delta_{\mathrm{CP}} = 0$  and $\delta_{\mathrm{CP}} = \pm\pi/2$  spectra in both neutrino and antineutrino modes, reflecting the sensitivity of the NO$\nu$A configuration to the CP-violating phase through the appearance channel. 

The information relevant for the estimation of the leptonic CP phase is fundamentally encoded in the reconstructed spectral information of these experiments. The dependence of the event spectra shown in Fig.~\ref{fig:t2k_spectrum} and Fig.~\ref{fig:nova_spectrum} on $\delta_{\rm CP}$ originates directly from the underlying oscillation probabilities. The spectral separation between different $\delta_{\rm CP}$ hypotheses becomes more pronounced near the first oscillation maximum, where the CP-sensitive interference terms are significant. Furthermore, since the matter potential enters with opposite sign for neutrino and antineutrino propagation, distinct differences are observed between the two beam modes for both NO and IO. These effects are more prominent in NO$\nu$A because of its longer baseline and consequently stronger matter effects compared to T2K.

\section{Results on Estimation of $\delta_{CP}$}
\label{sec:results}
Accelerator-based LBL experiments provide a controlled source of neutrinos and can operate in both neutrino and antineutrino beam modes. Since neutrinos and antineutrinos interact differently with the Earth's matter during propagation, the corresponding oscillation probabilities are modified differently by matter effects. Therefore, in order to isolate genuine leptonic CP-violating effects from matter-induced asymmetries, precise measurements of the $\nu_{\mu} \to \nu_e$ and $\bar{\nu}_{\mu} \to \bar{\nu}_e$ appearance channels as a function of energy are required.

The evolution of neutrino and antineutrino states  differs at the fundamental level due to the opposite sign of the matter potential and the transformation $\delta_{\rm CP}\to -\delta_{\rm CP}$. Consequently, the QFI associated with $\delta_{\rm CP}$ is expected to differ between neutrino and antineutrino modes, as well as for NO and IO. Since the expected event spectra and statistics are different for NO$\nu$A and T2K in both beam modes, it is important to analyze the neutrino and antineutrino channels separately in the subsequent discussion.

\subsection{QFI for neutrino and antineutrino state}
\label{subsec:asym}
\begin{figure*}[htb]
    \centering
    \includegraphics[width=1\linewidth]{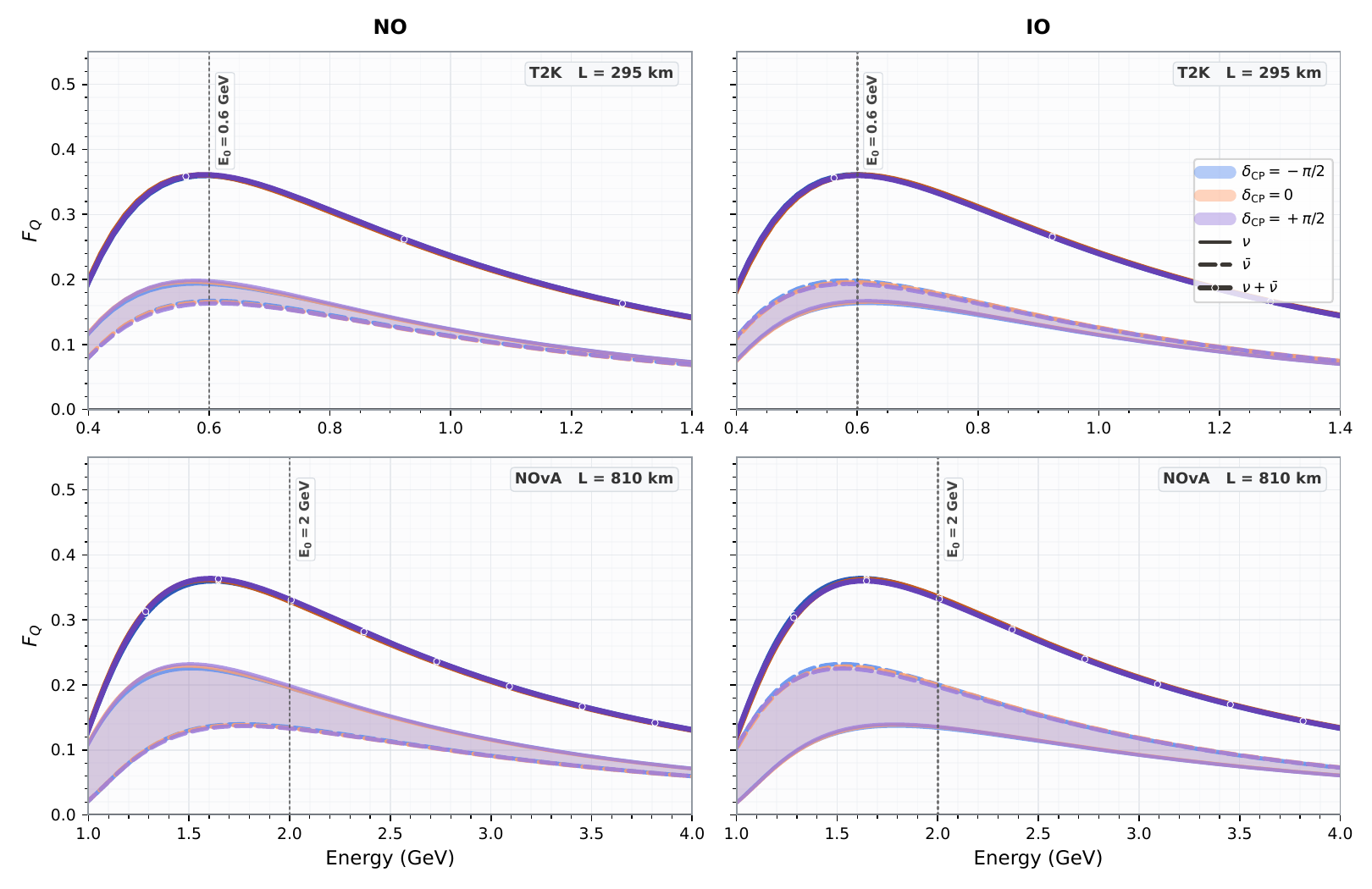}
    \caption{QFI associated with $\delta_{\rm CP}$ as a function of neutrino energy for T2K and NO$\nu$A for both neutrino and antineutrino states. The left panel corresponds to NO, while the right panel shows the results for IO. Vertical dashed lines denote flux peak energies. See the text for a detailed explanation.}
    \label{fig:qfi}
\end{figure*}

In Fig.\ref{fig:qfi}, we demonstrate the behavior of QFI for the leptonic CP violating phase as a function of energy for T2K and NO$\nu$A for both states i.e. neutrino and antineutrino. The QFI associated with both states shows a broader maxima around the first oscillation maximum and gradually decreases with the increase in energy for both T2K and NO$\nu$A. We show the values of QFI for $\delta_{\rm CP} = -\pi/2,\, 0, \,\pi/2$, and the dependence on the $\delta_{\rm CP}$ is evident to be weaker indicating that the change in the neutrino state associated with the change in the CP phase remains nearly stable. 

On the other hand, an asymmetry is observed between the neutrino and antineutrino state if we change the neutrino mass ordering due to the sign flip of the matter potential for neutrino and antineutrino evolution in matter. The observed QFI is larger in the neutrino state for NO as compared to the antineutrino state in the same ordering. For IO, the QFI is found to be larger in magnitude for antineutrino than the neutrino state. We observe the same behavior in both T2K and NO$\nu$A, but the difference is evident to be larger in NO$\nu$A, owing to the strong matter effects due to longer baseline.

This result suggests that the amount of information encoded about the parameter $\delta_{\rm CP}$, is intrinsically different for neutrino and antineutrino quantum state. This feature vanishes when we combine both states and the resultant combined QFI is same for both NO and IO and is almost independent of the chosen values of $\delta_{\rm CP}$. The magnitude also remains same for T2K and NO$\nu$A both, as the matter effects also gets canceled in the combination of both states. Hence, it becomes important to consider the neutrino and antineutrino mode separately in the analysis, since the corresponding event spectra in these LBL experiments differ significantly. This also makes the comparison between the QFI and the event-level Fisher information particularly relevant for understanding the CP sensitivity of T2K and NO$\nu$A from information perspective.

To further investigate the disparity in QFI for NO and IO shown in Fig.~\ref{fig:qfi}, we define an asymmetry, $A_{\text{QFI}}$, between the calculated values of QFI in neutrino and antineutrino state which measures the relative difference between them:

\begin{equation}
 A_{\text{QFI}} = \frac{F_Q^{\nu} - F_Q^{\bar{\nu}}}{F_Q^{\nu} + F_Q^{\bar{\nu}}}\;, 
\end{equation}
where $F_Q^{\nu}$ and $F_Q^{\bar{\nu}}$ corresponds to the QFI associated with neutrino and antineutrino state, respectively. Furthermore, to gain more insights into the behavior of $A_{\text{QFI}}$ , we also consider the probability asymmetry, defined as,
\begin{equation}
A_{\rm CP} = \frac{P(\nu_\mu \to \nu_e) - P(\bar{\nu}_\mu \to \bar{\nu}_e)}{P(\nu_\mu \to \nu_e) + P(\bar{\nu}_\mu \to \bar{\nu}_e)} \;.
\label{eq:ACP_def}
\end{equation}

\begin{figure}[H]
    \centering
    \includegraphics[width=1\linewidth]{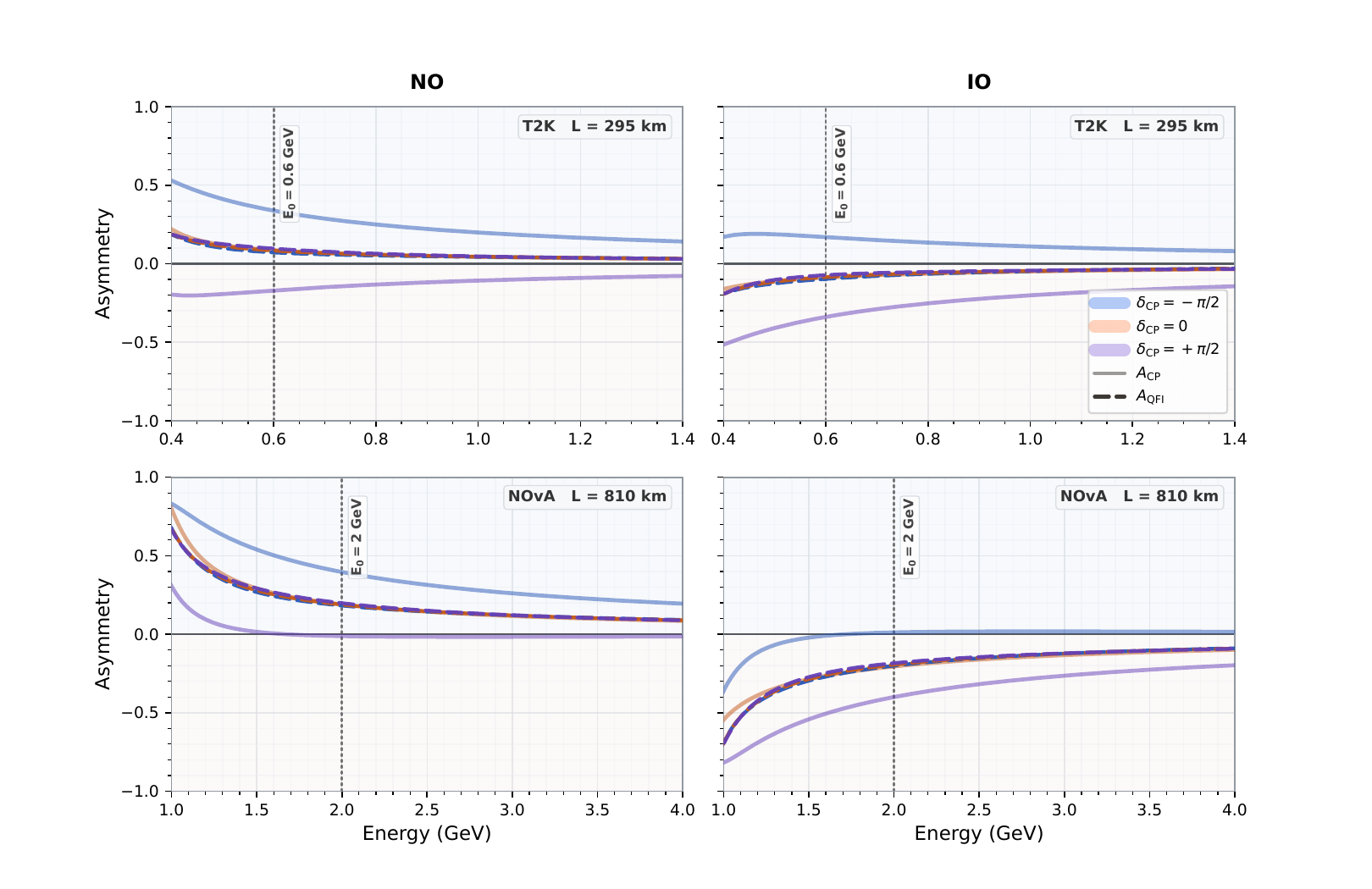}
    \caption{The probability asymmetry $A_{\rm CP}$ (solid) and QFI asymmetry $A_{\text{QFI}}$ (dashed) as a function of the neutrino energy for T2K (upper panels) and NO$\nu$A (lower panels) setups. The left panels consider the case of NO while right panels represent the IO scenario. See the text for a detailed explanation.}
    \label{fig:acp}
\end{figure}

While \(A_{\rm CP}\) characterizes the magnitude and sign of CP violation at the oscillation probability level, \(A_{\text{QFI}}\) quantifies the relative amount of quantum information about the CP-violating phase \(\delta_{\rm CP}\) carried by neutrino and antineutrino states. In Fig.~\ref{fig:acp}, we show the probability asymmetry \(A_{\rm CP}\) and the QFI asymmetry \(A_{\text{QFI}}\) as functions of neutrino energy for the T2K (upper panels) and NO$\nu$A (lower panels) configurations at three representative values of the Dirac CP phase, \(\delta_{\rm CP} = -\pi/2,\,0,\,\pi/2\). 

For the NO scenario in NO$\nu$A, both asymmetries, \(A_{\rm CP} \, \text{and} \, A_{\text{QFI}}\), remain positive over the considered energy range for all benchmark values of \(\delta_{\rm CP}\) as the matter potential enhances \(\nu_\mu \to \nu_e\) transition probability. In contrast, for IO, the asymmetries exhibit the opposite behavior as the corresponding antineutrino channel is enhanced due to matter effects. This behavior indicates the matter enhancement in the probabilities is ordering dependent. The same behavior is also observed for QFI asymmetry between both quantum states.

For the T2K configuration with NO, both asymmetries are also predominantly positive, except for the case \(\delta_{\rm CP}=+\pi/2\) (purple solid curve). On the other hand, in the IO scenario, both asymmetries become negative over most of the energy range, except for the case \(\delta_{\rm CP}=-\pi/2\) (blue solid curve). The variation in the sign of the asymmetries in T2K arises from the combined effect of CP violation and matter effects in neutrino and antineutrino propagation. The behavior of $A_{\text{QFI}}$ demonstrates that the intrinsic quantum sensitivity to $\delta_{\rm CP}$ already contains a hierarchy-dependent asymmetry between neutrino and antineutrino propagation. 

Despite the clear sensitivity of \(A_{\text{QFI}}\) to the neutrino mass ordering for both T2K and NO$\nu$A, its dependence on the value of \(\delta_{\rm CP}\) remains relatively weak within a given ordering. This indicates that although the QFI asymmetry efficiently captures the hierarchy-dependent differences between neutrino and antineutrino propagation, it is comparatively less sensitive to variations in the CP phase itself. It should also be noted that a large probability asymmetry may also imply a significant variation in the intrinsic quantum sensitivity encoded in the propagated neutrino state.

It is important to emphasize that the QFI itself is not a directly experimentally observable quantity, but rather represents the ultimate quantum limit on parameter estimation achievable over all possible measurements. In LBL experiments, the experimentally accessible observables are reconstructed neutrino and antineutrino event spectra, which involve different event statistics. Therefore, using the additivity property of QFI, the relevant comparison is between the event-level Fisher information and the event-weighted QFI benchmark, rather than the intrinsic single-neutrino QFI itself. Motivated by this observation, in the following subsections we investigate the detector-level extraction of CP information through the event-weighted QFI and the corresponding event-level Fisher information constructed from neutrino spectra.

\subsection{Event-weighted Quantum Fisher Information}

The QFI discussed in the previous subsection represents the intrinsic quantum sensitivity associated with a single evolved neutrino state. In the following analysis, we instead consider the event-weighted QFI defined in Eq.~\eqref{eq:wqfi}. This quantity serves as an effective benchmark for the intrinsic quantum sensitivity of neutrino oscillation states while simultaneously incorporating the event-level statistical information accessible in the experiments.

\begin{figure}[H]
    \centering
    \includegraphics[width=1\linewidth]{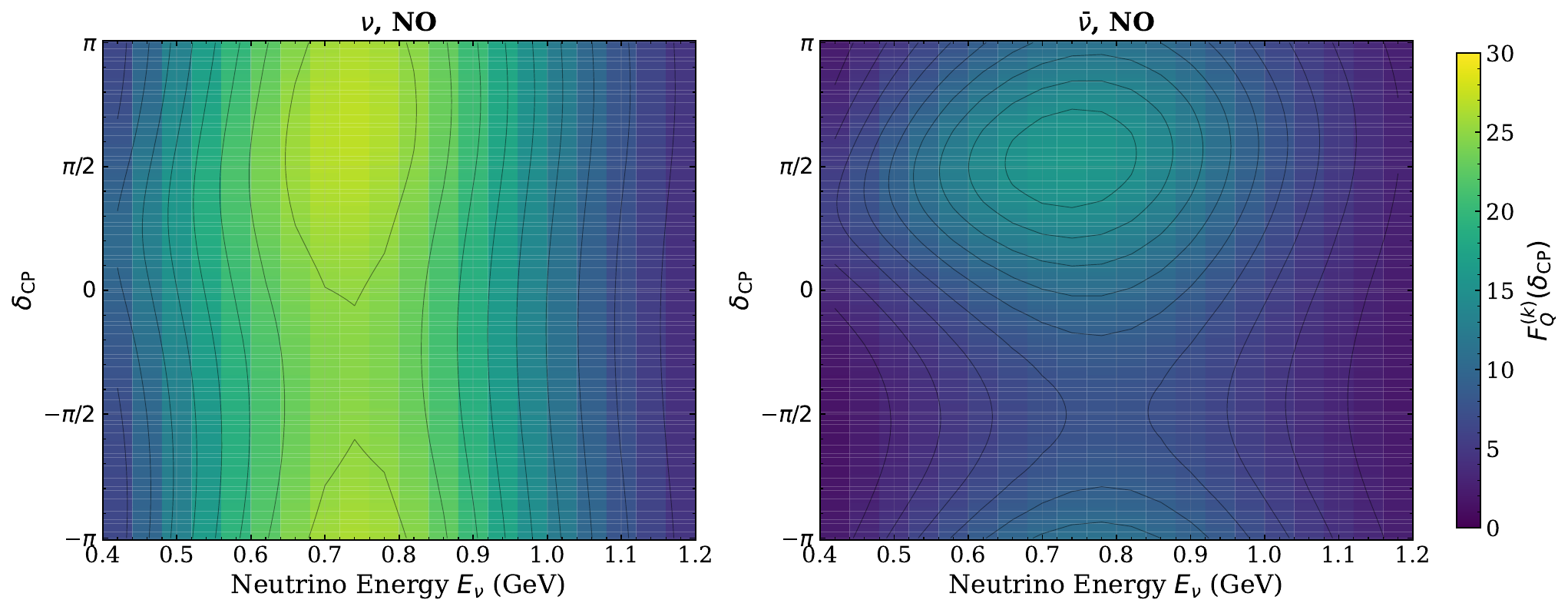}\\
    \includegraphics[width=1\linewidth]{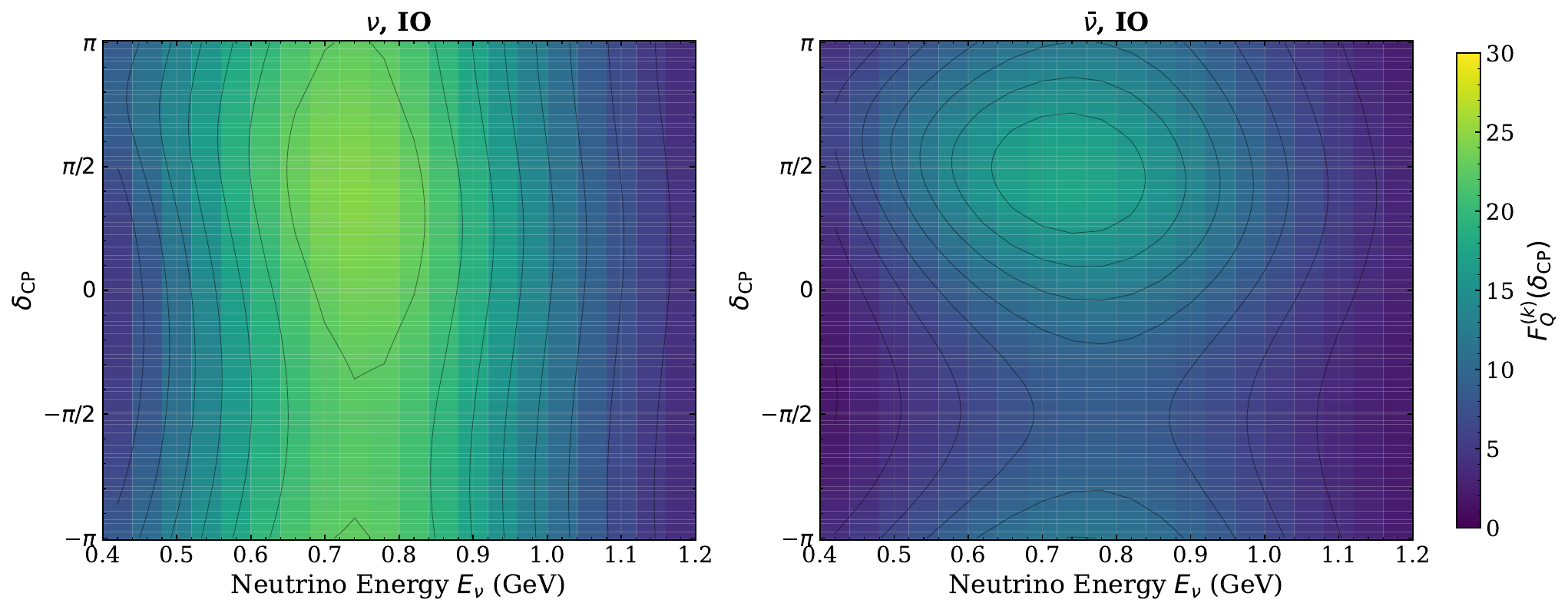}
    \caption{Event-weighted QFI $F^{(k)}_{Q}(\delta_{\rm CP})$ as a function of the neutrino energy $E_\nu$ and the leptonic CP-violating phase $\delta_{\rm CP}$ for the T2K setup. The left panels correspond to the neutrino mode, while the right panels show the antineutrino mode. The upper and lower panels assume NO and IO, respectively. See the text for a detailed explanation.}
    \label{fig:t2kwqfi}
\end{figure}

We show the event-weighted QFI as a function of neutrino energy $E_\nu$ and the leptonic CP phase $\delta_{\rm CP}$ for the T2K configuration in Fig.~\ref{fig:t2kwqfi}. The left and right panels correspond to the neutrino and antineutrino modes, respectively, while the upper and lower panels assume NO and IO. It is apparent from the figure that the event-weighted QFI exhibits only a mild dependence on the value of $\delta_{\rm CP}$, in contrast to the stronger variation observed for the intrinsic QFI of the propagated quantum state. This behavior arises because the weighting factor $C_k$ depends only on the unoscillated event rates. The difference between the neutrino and antineutrino modes primarily reflects the higher event statistics in the neutrino beam mode. For both beam modes, the event-weighted QFI is largest around the peak neutrino energy. Owing to the relatively short baseline and the small matter effects in T2K, the difference between the normal and inverted mass orderings remains marginal. 

\begin{figure}[htb!]
    \centering
    \includegraphics[width=1\linewidth]{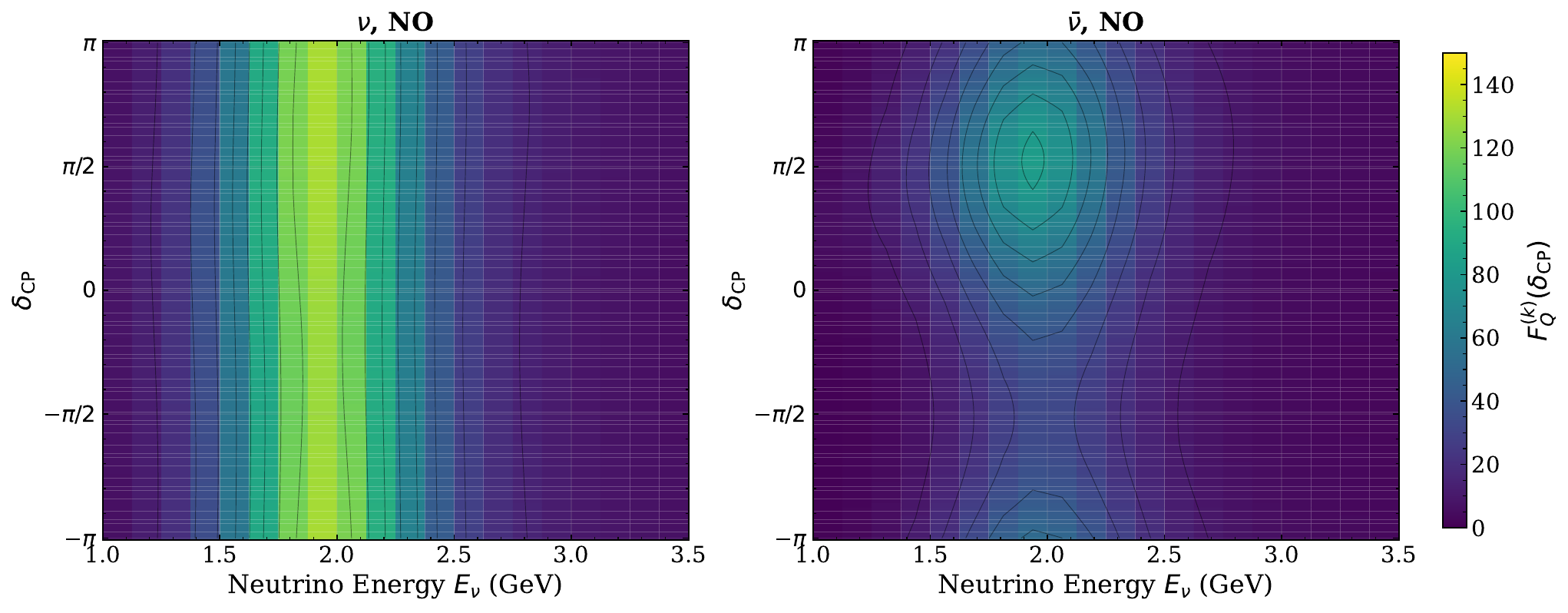}\\
    \includegraphics[width=1\linewidth]{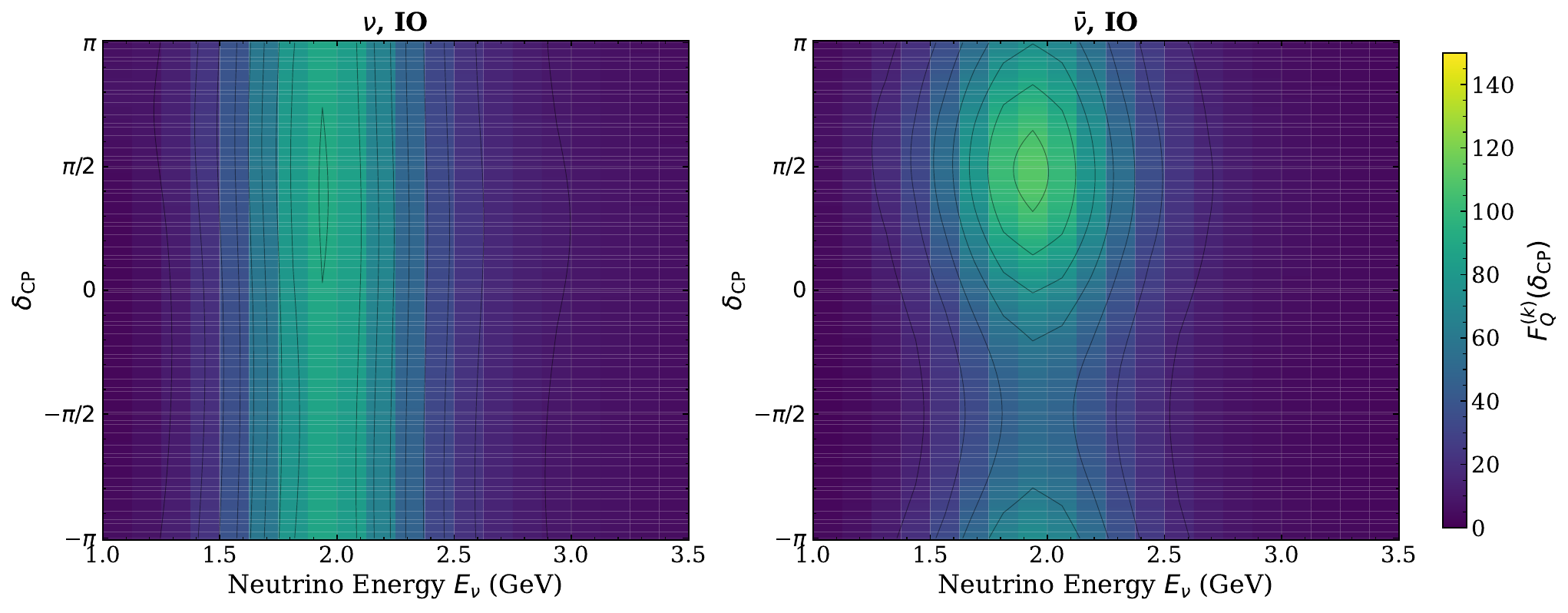}
    \caption{Event-weighted QFI $F^{(k)}_{Q}(\delta_{\rm CP})$ as a function of the neutrino energy $E_\nu$ and the leptonic CP-violating phase $\delta_{\rm CP}$ for the NO$\nu$A configuration. The left panels correspond to the neutrino mode, while the right panels show the antineutrino mode. The upper and lower panels assume NO and IO, accordingly. See the text for a detailed explanation.}
    \label{fig:novawqfi}
\end{figure}

In Fig.~\ref{fig:novawqfi}, we present the corresponding event-weighted QFI distributions for the NO$\nu$A configuration. The left panels correspond to the neutrino mode, while the right panels show the antineutrino mode. The upper and lower panels assume NO and IO, respectively. As in the case of T2K, the overall distribution for NO$\nu$A remains qualitatively similar, reflecting the similar structure of the propagated quantum state. Nevertheless, owing to its longer baseline and enhanced matter effects, NO$\nu$A exhibits a more pronounced dependence on the neutrino mass ordering and the beam mode.

\subsection{Event-level Fisher Information}
The experimentally accessible sensitivity to $\delta_{\rm CP}$ is governed by the variation of the reconstructed event spectra with respect to the CP phase. The corresponding change in the expected number of events in each reconstructed energy bin can be written using Eq.~\eqref{eq:event_rate} as,
\begin{equation}
\frac{\partial N_k}{\partial \delta_{\rm CP}}
=
\int dE_\nu \,
\Phi(E_\nu)\,
\sigma(E_\nu)\,
\epsilon(E_\nu)\,
R(E_\nu,E_k)\,
\frac{\partial P_{\mu e}(E_\nu,\delta_{\rm CP})}
{\partial \delta_{\rm CP}}.
\label{eq:diffevent}
\end{equation}
This quantity can be used directly to determine the event-level Fisher information associated with the event spectra per energy bin. Similar to the flavor POVM Fisher information, the experimentally accessible sensitivity is ultimately governed by the variation of the oscillation probability with respect to $\delta_{\rm CP}$. However, in experiments this information is embedded within reconstructed event spectra after including flux, cross-section, and detector response. Hence, by using the quantity from Eq.~\eqref{eq:diffevent} in Eq.~\eqref{eq:Fevent}, the event-level Fisher information associated with the experimentally reconstructed spectra can be obtained.

\begin{figure}[H]
    \centering
    \includegraphics[width=1\linewidth]{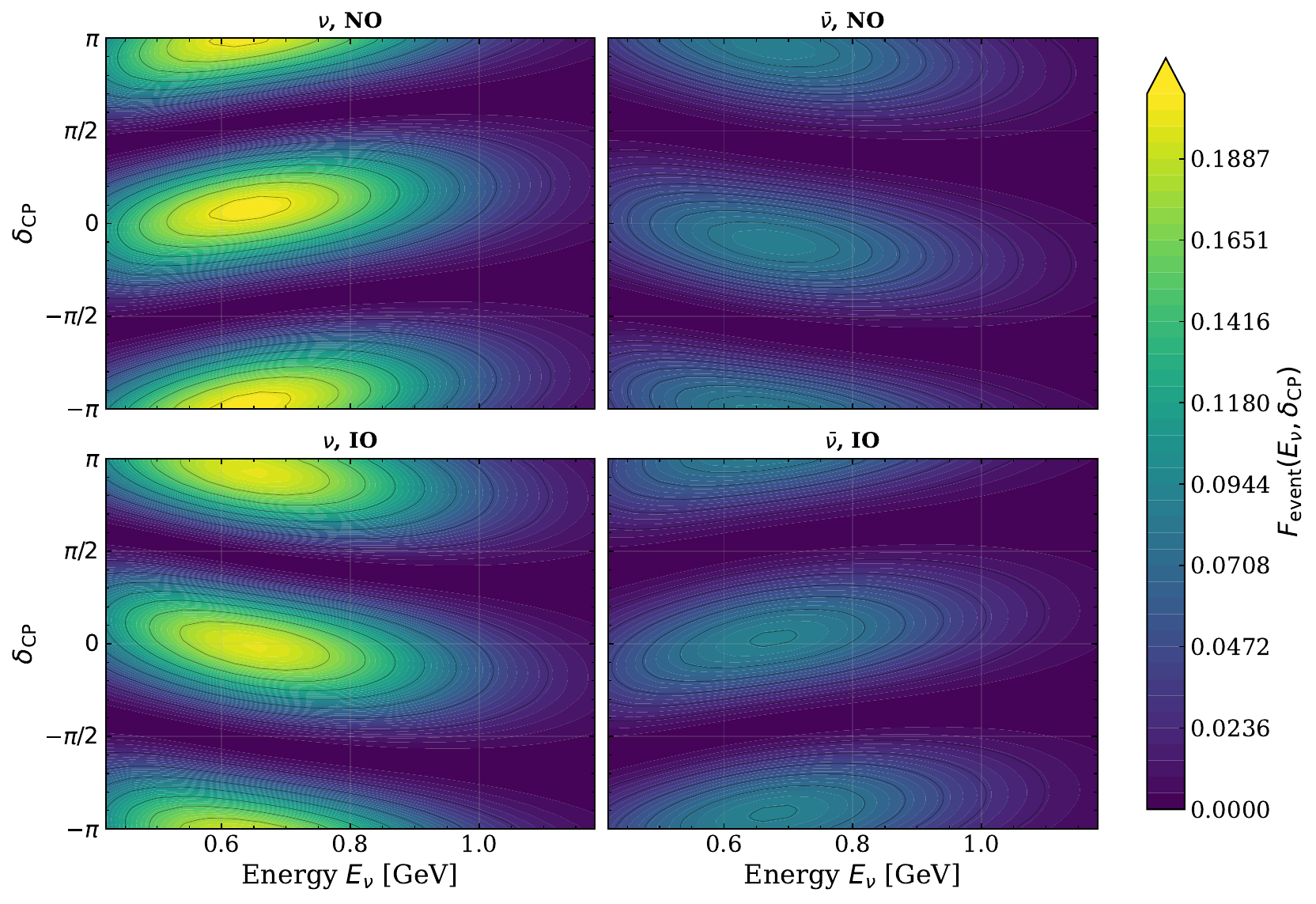}
    \caption{Event-level Fisher information $F_{\rm event}(E_\nu,\delta_{\mathrm{CP}})$ for the T2K configuration. The color scale represents the Fisher information contribution per reconstructed energy bin as a function of $\delta_{\mathrm{CP}}$ (vertical axis) and neutrino energy $E_\nu$ (horizontal axis). The left (right) panels show the neutrino (antineutrino) mode, while the upper (lower) rows assume NO (IO). See the text for a detailed explanation.}
    \label{fig:t2kevent}
\end{figure}

Fig.~\ref{fig:t2kevent} presents the event-level Fisher information distributions for the T2K configuration. The dominant information region ($F_{\rm event}(E_\nu,\delta_{\rm CP}) \sim 0.2$) appears near the first oscillation maximum around $E_\nu\sim0.6$~GeV and close to the CP-conserving phases $\delta_{\rm CP}=0$ and $\pm\pi$. Due to the shorter baseline and consequently weaker matter effects, the neutrino and antineutrino modes exhibit comparatively similar overall structures with lower values of $F_{\rm event}(E_\nu,\delta_{\rm CP})$ for the case of antineutrino mode due to lower statistics, although the contour tilts are reversed between the two beam modes. This opposite tilt originates from the CP-conjugate interference structure in the oscillation probabilities through the replacement $\delta_{\rm CP}\to-\delta_{\rm CP}$ for antineutrinos.

The distinction between NO and IO remains relatively weak in both magnitude and overall distribution, reflecting the limited hierarchy sensitivity of T2K. The event-level Fisher information is minimized near the maximally CP-violating values $\delta_{\rm CP}=\pm\pi/2$, where the local variation of the reconstructed spectra with respect to $\delta_{\rm CP}$ becomes comparatively weaker. Compared to the corresponding event-weighted QFI distributions in Fig.~\ref{fig:t2kwqfi}, the event-level Fisher information remains smaller, reflecting the information loss associated with the flavor measurement scheme.

\begin{figure}[H]
    \centering
    \includegraphics[width=1\linewidth]{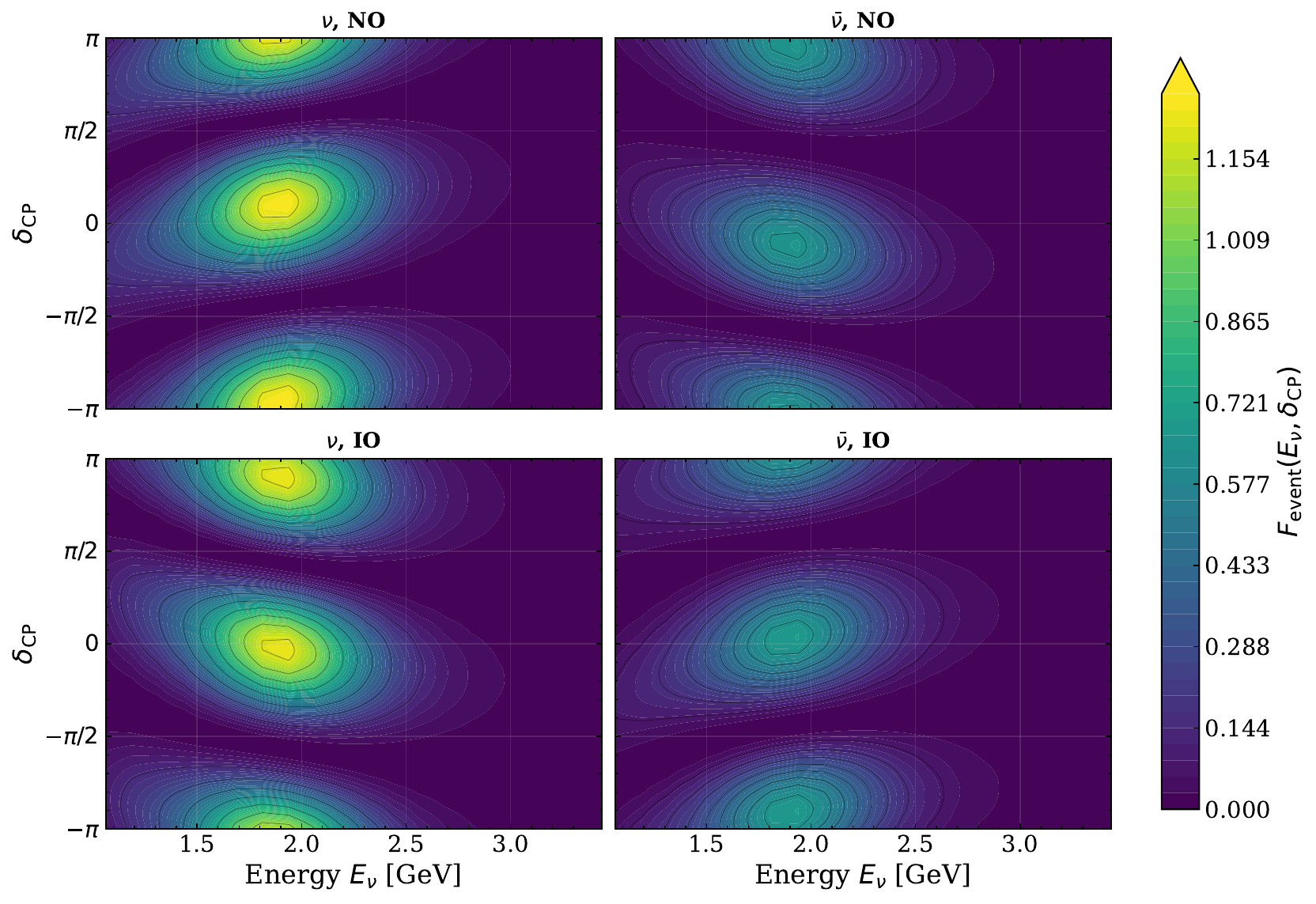}\\
    \caption{Event-level Fisher information $F_{\rm event}(E_\nu,\delta_{\mathrm{CP}})$ for the NO$\nu$A configuration. The color scale represents the Fisher information contribution per reconstructed energy bin as a function of the CP-violating phase $\delta_{\mathrm{CP}}$ (vertical axis) and neutrino energy $E_\nu$ (horizontal axis). The left (right) panels show the neutrino (antineutrino) mode, while the upper (lower) rows assume NO (IO). See the text for a detailed explanation}
    \label{fig:novaevent}
\end{figure}

Fig.~\ref{fig:novaevent} shows the corresponding event-level Fisher information distributions for the NO$\nu$A configuration. For both orderings, the neutrino mode exhibits the dominant information region ($F_{\rm event}(E_\nu,\delta_{\rm CP}) \sim 1.2$) near the first oscillation maximum around $E_\nu \sim 2$~GeV, particularly close to the CP-conserving phases $\delta_{\rm CP}=0$ and $\pm\pi$, while the antineutrino mode carries comparatively smaller information. Unlike the T2K configuration, NO$\nu$A exhibits a comparatively narrower information distribution due to its stronger matter effects and more localized energy spectrum. Furthermore, the event-level Fisher information $F_{\rm event}(E_\nu,\delta_{\rm CP})$ shows no strong hierarchy dependence in the overall contour structure for both neutrino and antineutrino modes, although the relative magnitude of the information content is redistributed between the two beam modes.

Thus, we observe that the event-level Fisher information as shown in Figs.~\ref{fig:t2kevent} and \ref{fig:novaevent} does not fully saturate the corresponding event-weighted QFI as in Figs.~\ref{fig:t2kwqfi} and \ref{fig:novawqfi}. Therefore, the comparison between these two quantities becomes important for analyzing the efficiency of these experiments in extracting the information available for the estimation of the parameter under consideration.

\subsection{Experimental Extraction Efficiency}

\begin{figure}[H]
    \centering
    \includegraphics[width=0.9\linewidth]{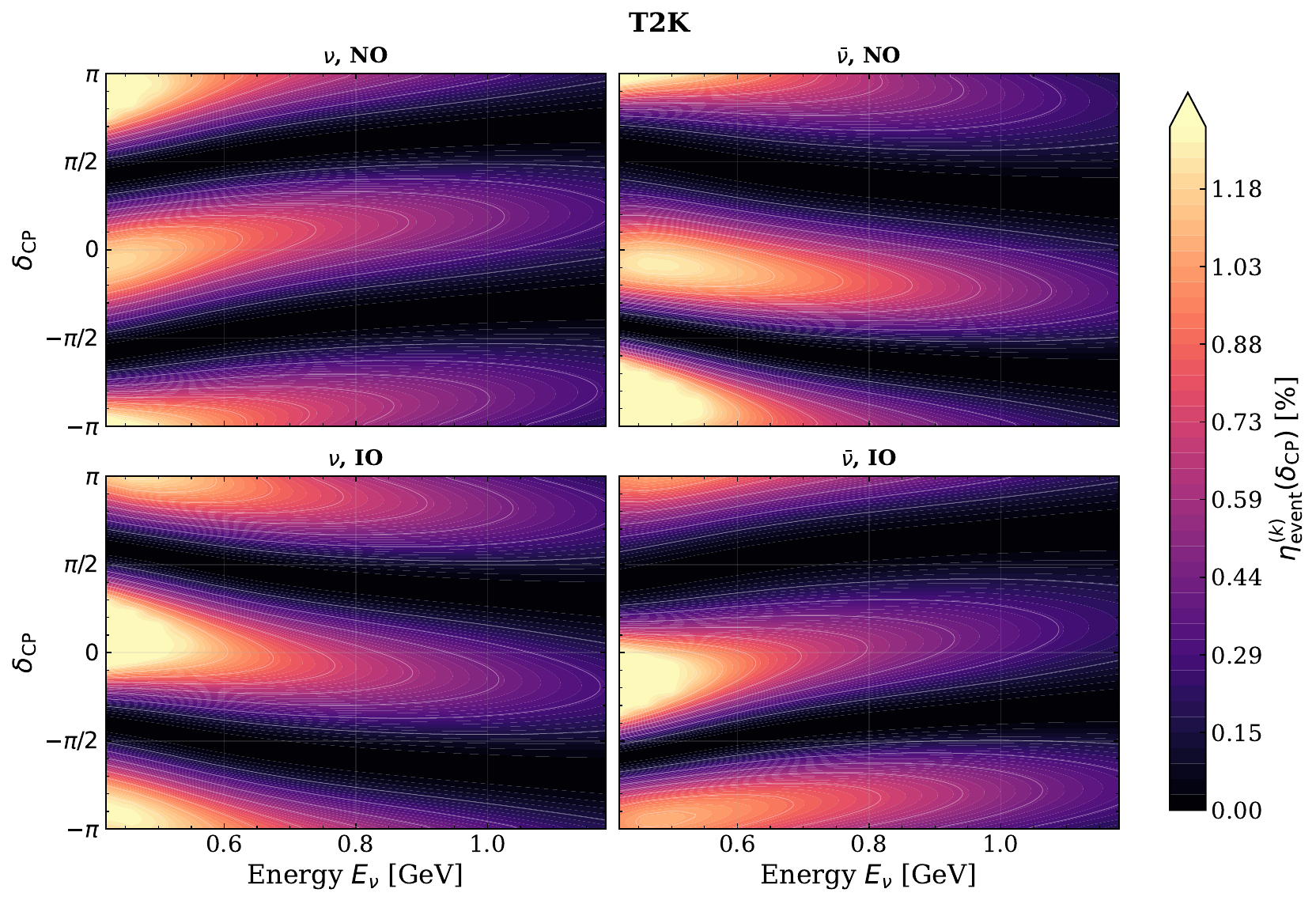}\\
    \includegraphics[width=0.9\linewidth]{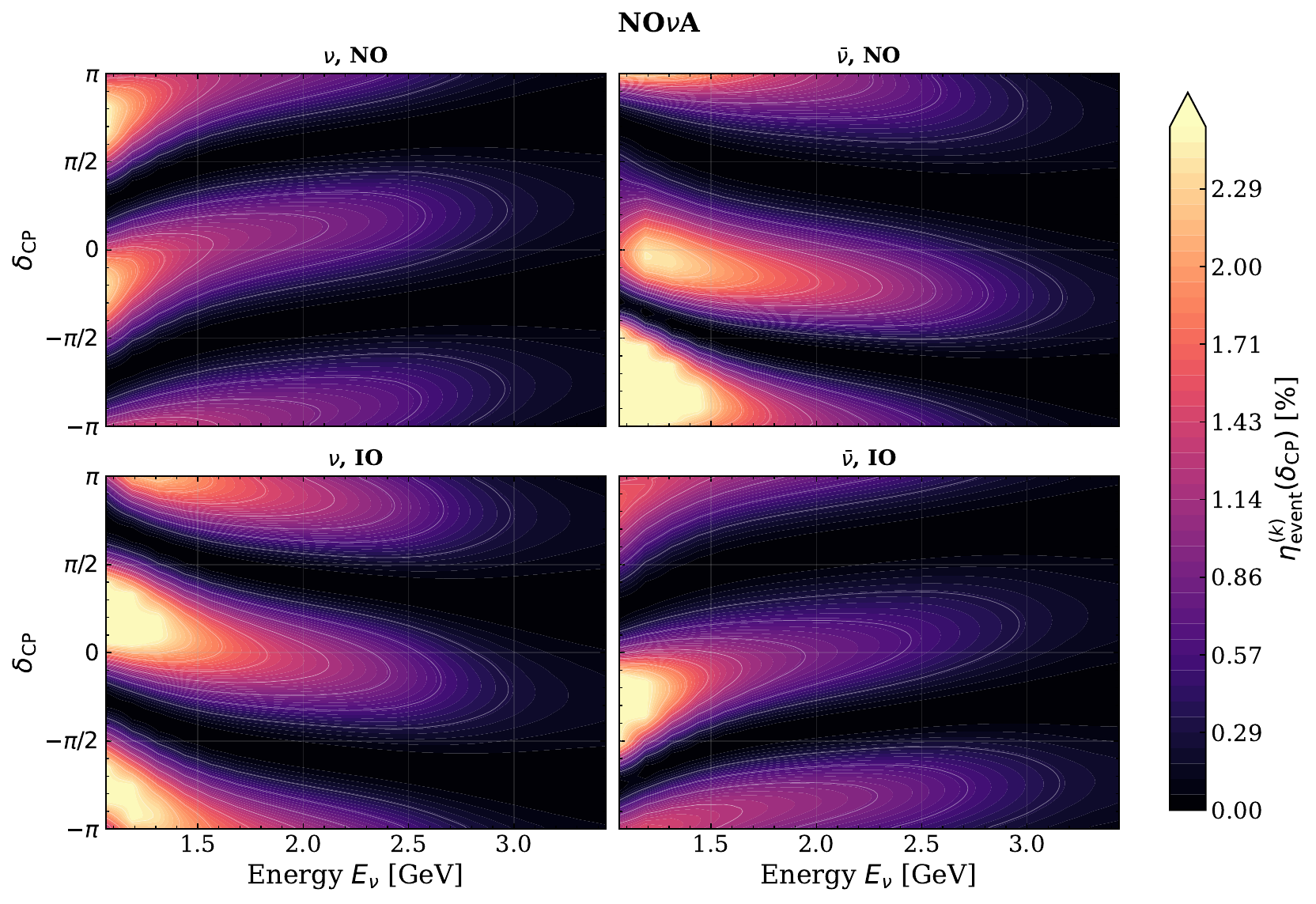}
    \caption{Extraction efficiency $\eta_{\rm event}^{(k)}(\delta_{\mathrm{CP}})$ for the T2K and NO$\nu$A configurations. The color scale (0 to 1) indicates the fraction of the extraction efficiency that is accessible via the experimental measurement. The horizontal axis represents the neutrino energy $E_\nu$, and the vertical axis refers to $\delta_{\mathrm{CP}}$. Panels show neutrino (left) and antineutrino (right) modes, with NO (top) and IO (bottom). See the text for a detailed explanation.}
    \label{fig:eta}
\end{figure}

Hence, in order to quantify how efficiently the experimentally reconstructed spectra extract the available quantum information associated with $\delta_{\rm CP}$, we define a ratio:

\begin{equation}
\eta_{\rm event}^{(k)}(\delta_{\rm CP})
=
\frac{
F_{\rm event}^{(k)}(\delta_{\rm CP})
}{
F_Q^{(k)}(\delta_{\rm CP})
}\;.
\end{equation}
This quantity illustrates the associated fraction of the event-weighted QFI benchmark that is accessible after the inclusion of detector effects. In Fig.~\ref{fig:eta} we show the behavior of the bin-wise extraction efficiency as a function of neutrino energy $E_\nu$ (horizontal axis) and the leptonic CP phase $\delta_{\rm CP}$ (vertical axis) for both the T2K and NO$\nu$A configurations. The brighter regions in the contour represent the ranges of energy and CP phase where the amount of extracted information from the effective event-weighted QFI in the detectors is maximal, while the darker regions correspond to limited extraction efficiency. For T2K, the extraction efficiency remains at the sub-percent and percent level, over most of the parameter space. This behavior follows a similar pattern to that of $F_{\rm event}^{(k)}$ for both neutrino and antineutrino modes and shows weak hierarchy dependence between the two mass orderings. It is evident from the figure that around $\delta_{\rm CP}=\pm\pi/2$, the extraction efficiency becomes very low irrespective of the presence of information in the quantum state.

For NO$\nu$A, the extraction efficiency reaches slightly larger values than in T2K, up to $\sim 2\%$ in the low energy bins. The overall structure of the distribution is comparatively less broad than that of T2K, and the dependence on the neutrino mass ordering is more pronounced. For NO, the antineutrino mode shows comparatively larger extraction efficiency than the neutrino mode, while for IO this behavior reverses. At lower energies, NO$\nu$A exhibits better extraction efficiency near the CP-violating values of $\delta_{\rm CP}$ as compared to T2K. However, as the energy increases, the extraction efficiency gradually decreases toward smaller values.

\begin{figure}[H]
    \centering
    \includegraphics[width=0.9\linewidth]{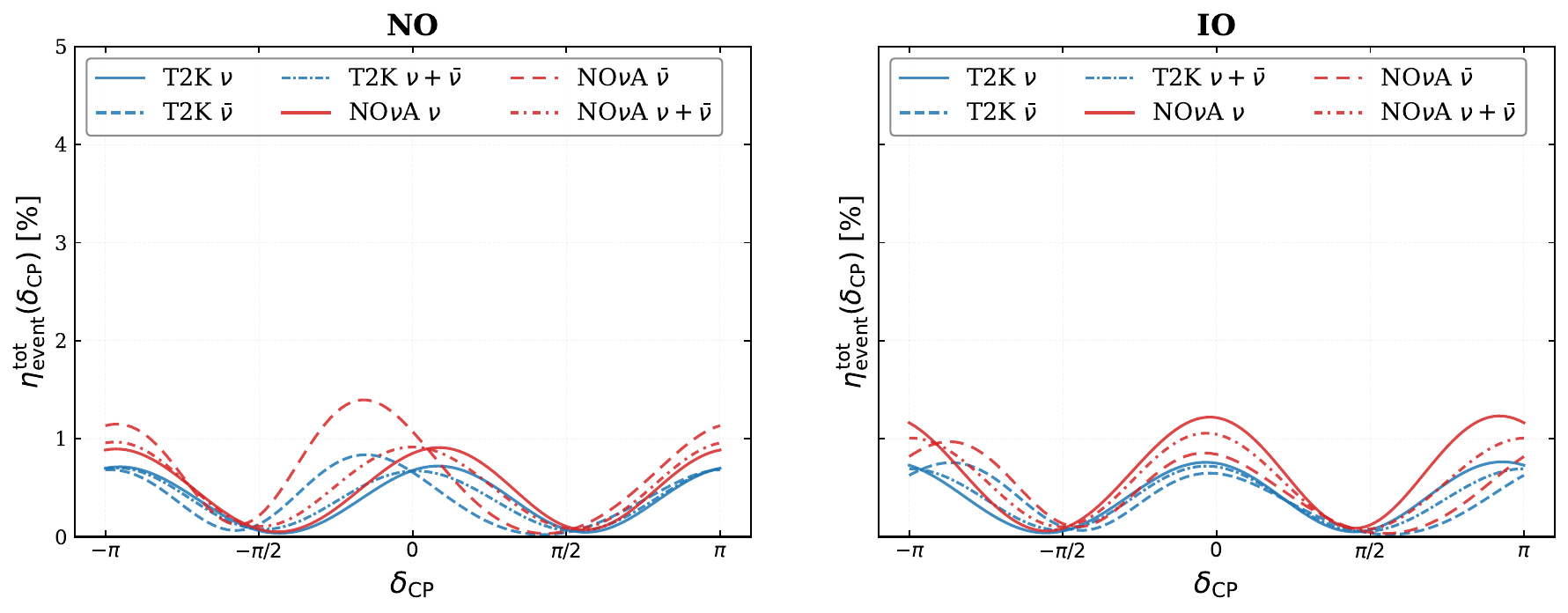}
    \caption{Total extraction efficiency $\eta_{\rm event}^{\rm tot}(\delta_{\rm CP})$ for the T2K (blue) and NO$\nu$A (red) configurations. For both experiments, solid lines represent the neutrino mode, dashed lines the antineutrino mode, and dot-dashed lines the combined neutrino plus antineutrino modes. See the text for a detailed explanation.}
    \label{fig:toteta}
\end{figure}

In Fig.~\ref{fig:toteta} we present the total extraction efficiency, defined as, 
\begin{equation}
\eta_{\rm event}^{\rm tot}(\delta_{\rm CP})
=
\frac{
F_{\rm event}^{\rm tot}(\delta_{\rm CP})
}{
F_Q^{\rm tot}(\delta_{\rm CP})
},
\end{equation}
for the T2K (blue) and NO$\nu$A (red) experimental setups. In both experiments, the solid, dashed, and dot-dashed lines correspond to the neutrino mode, antineutrino mode, and the combination of both modes, respectively. It is evident from the figure that for both experiments, the total extraction efficiency remains significantly below the event-weighted quantum benchmark for both beam modes. This shows that these experiments do not fully access the intrinsic quantum information available in the neutrino state for the estimation of $\delta_{\rm CP}$.

For both T2K and NO$\nu$A configurations, the total extraction efficiencies remain at the sub-percent $\mathcal{O}(1\%)$ level. For the case of the NO$\nu$A under IO, the total extraction efficiency is slightly larger than that of T2K for most of the $\delta_{\rm CP}$ values while for NO it remains almost similar. This trend is partially due to the fact that the NO$\nu$A configuration experiences larger matter effects and enhanced flavor asymmetry as compared to T2K, as we discussed in Section~\ref{subsec:asym}. It is to be noted that the extraction efficiency almost vanishes around the maximal CP-violating values of the CP phase i.e. $\delta_{\rm CP}\simeq \pm\pi/2$ for both experiments and all beam modes, indicating that event spectra become locally indistinguishable under small CP-phase variations regardless of the intrinsic quantum information.  This demonstrates that these LBL experiments may access only a tiny fraction of intrinsic quantum information available in the state for the CP phase.

Under the assumption of IO, the T2K setup shows no significant change with the reversal of neutrino mass ordering and behaves similarly to the NO case. In contrast, for NO$\nu$A, although the magnitude of the maximal extraction efficiency remains nearly unchanged, the dominant contribution shifts from the antineutrino mode to the neutrino mode. Upon combining both beam modes, the maximal extraction efficiency again becomes similar to that observed for NO. When combining the Fisher information for both neutrino and antineutrino mode, the overall statistical information indeed gets slightly improved but it washes out the feature of asymmetry observed in the separate beam modes.

\section{Comparison between the estimation of $\theta_{13}$ and
$\delta_{\rm CP}$} 
\label{sec:theta13}

\begin{figure}[H]
    \centering
\includegraphics[width=1\linewidth]{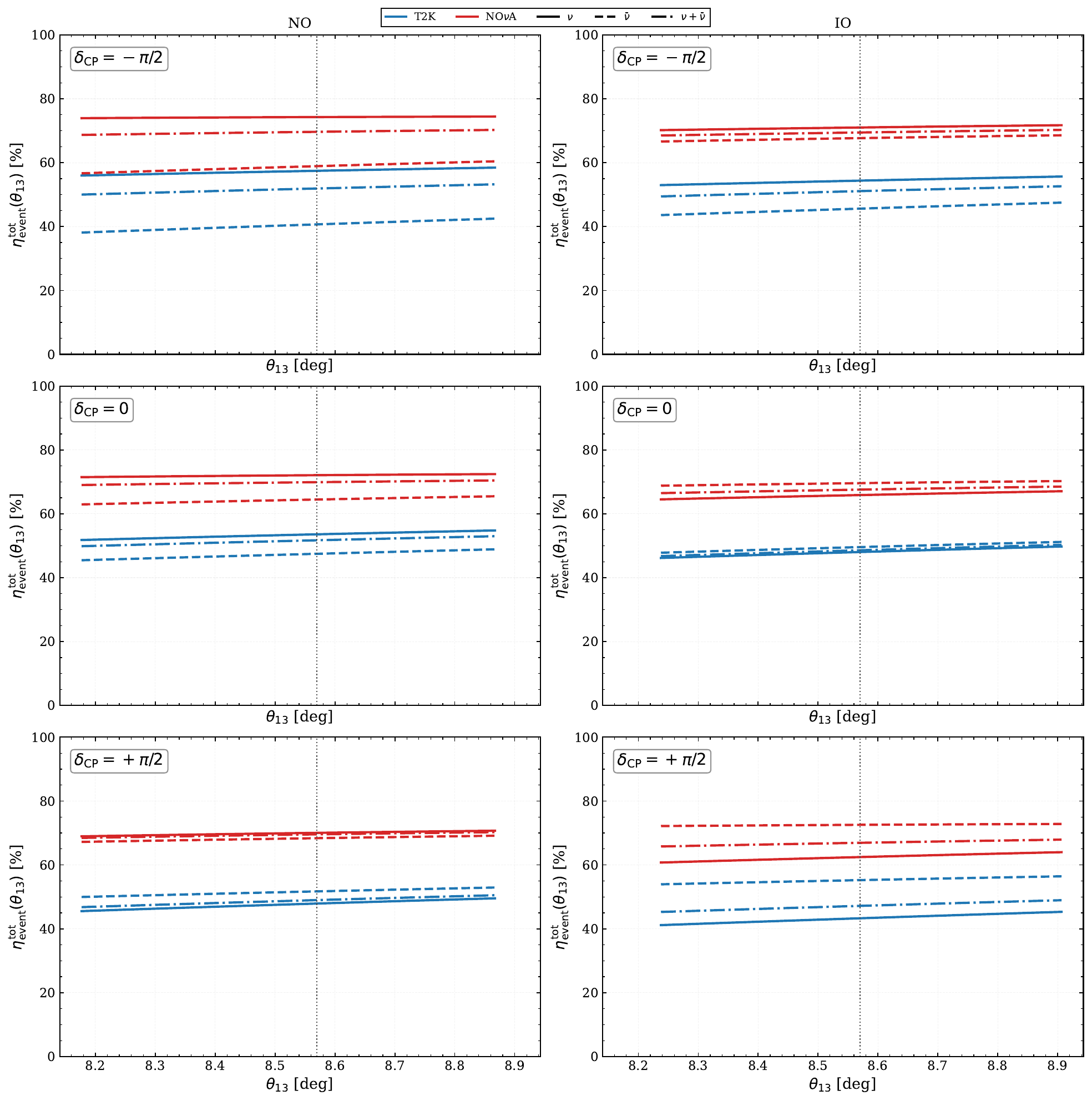}
\caption{Total extraction efficiency for estimating $\theta_{13}$ in T2K (blue) and NO$\nu$A (red). The left and right columns correspond to NO and IO, while the rows show
$\delta_{\rm CP}=-\pi/2$, $0$, and $+\pi/2$ from top to bottom. The solid, dashed, and dash-dotted curves represent the neutrino, antineutrino, and combined samples, respectively. The vertical dotted line represents the best-fit value $\theta_{13}=8.57^{\circ}$~\cite{Esteban:2024eli}. See the text for a detailed explanation.}
    \label{fig:eta_theta13}
\end{figure}

To validate the results obtained for the CP phase, we now apply the same framework described in our analysis to the reactor mixing angle $\theta_{13}$, which is one of the most precisely measured oscillation parameters. We perform this analysis for both the T2K and NO$\nu$A configurations. This comparison will also highlight that the quantum information accessible to an experiment depends on the oscillation parameter to be estimated. For instance, at the level of the quantum state, variations of $\theta_{13}$ modifies the flavor amplitudes whereas $\delta_{\rm CP}$ affects the propagation phase. However, the accessible information in the experiments is determined by the flavor transition probabilities, whose parameter dependence majorly governs the event-level Fisher information. This dependence can be understood by the approximate expression of appearance probability, 
\begin{equation}
P_{\mu e} \simeq P_{\rm atm} + P_{\rm sol} + P_{\rm int} \;, \end{equation} 
where the atmospheric term scales as $P_{\rm atm}\propto\sin^2 2\theta_{13}$, the solar term $P_{\rm sol}$, is independent of $\theta_{13}$ at this order and the interference contribution behaves as
\begin{equation}
P_{\rm int} \propto \alpha\sin 2\theta_{13} \cos\!\left(\Delta+\delta_{\rm CP}\right), ~~ \Delta=\frac{\Delta m_{31}^{2} L}{4E_\nu}, ~~\alpha = \frac{\Delta m_{21}^{2}}{\Delta m_{31}^{2}} \simeq 0.03\;.
\end{equation}
Therefore, $\theta_{13}$ enters both the dominant atmospheric amplitude and the sub-leading interference term, while $\delta_{\rm CP}$ appears only through the $\alpha$-suppressed interference contribution. As shown in Ref.~\cite{Frugiuele:2026yeq}, for the T2K configuration the flavor measurement is optimal for the estimation of $\theta_{13}$ and can saturate the corresponding QFI, in contrast to the case of $\delta_{\rm CP}$. In this section, we follow the methodology introduced in Sec.~\ref{sec:method}, replacing the parameter to be estimated by $\theta_{13}$ to calculate single neutrino QFI, the event weighted quantum benchmark. For each reconstructed-energy bin, the event-level Fisher information is obtained from the reconstructed appearance spectrum as

\begin{equation} F_{\rm event}^{(k)}(\theta_{13}) = \frac{1}{N_k(\theta_{13})} \left[ \frac{\partial N_k(\theta_{13})} {\partial\theta_{13}} \right]^2 \;.
\end{equation}
Figure~\ref{fig:eta_theta13} shows the total information extraction efficiency for the estimation of the mixing angle $\theta_{13}$ over its allowed $3\sigma$ range, considering the three benchmark values of the CP-violating phase, $\delta_{\rm CP}=-\pi/2$, $0$, and $+\pi/2$. Since the reactor mixing angle is already a precisely measured parameter, the extraction efficiency exhibits only a mild variation across the allowed $\theta_{13}$ range. In T2K, $\eta_{\theta_{13}}$ lies between approximately $35$--$55\%$, whereas for NO$\nu$A it reaches about $70\%$ for both the neutrino and antineutrino beam modes. The overall efficiency depends on the assumed value of $\delta_{\rm CP}$, leading to a gradual shift of the efficiency curves. For the combined beam mode, the extraction efficiency lies between the corresponding neutrino and antineutrino values since the additivity applies to the Fisher information rather than to the efficiencies themselves. In contrast to the sub-percent and percent-level extraction efficiencies obtained for the estimation of $\delta_{\rm CP}$, the larger efficiencies observed here indicate that both T2K and NO$\nu$A can access a substantial fraction of the quantum information associated with the reactor mixing angle.

\begin{table}[H]
\centering
\caption{Total event-level Fisher information ($F_{\rm event}^{\text {tot}}$), weighted QFI ($F_{Q}^{\text {tot}}$), and information-extraction efficiency $\eta^{\text {tot}}$ for the combined neutrino and antineutrino mode ($\nu+\bar{\nu}$), shown separately for the estimation of $\theta_{13}$ (columns 4--6) and $\delta_{\rm CP}$ (columns 7--9).}
\label{tab:merged_fisher}
\renewcommand{\arraystretch}{1.15}
\begin{tabular}{ccccccccc}
\toprule
& & & \multicolumn{3}{c}{ $\theta_{13}$} & \multicolumn{3}{c}{ $\delta_{\rm CP}$} \\
\cmidrule(lr){4-6} \cmidrule(lr){7-9}
Experiment & Ordering & $\delta_{\rm CP}$ & $F_{\rm event}^{\theta_{13}}$ & $F_{Q}^{\theta_{13}}$ & $\eta_{\theta_{13}}^{\text {tot}}~(\%)$ & $F_{\rm event}^{\delta_{\rm CP}}$ & $F_{Q}^{\delta_{\rm CP}}$ & $\eta_{\delta_{\rm CP}}^{\text {tot}}~(\%)$ \\
\midrule

\multirow{3}{*}{T2K}
& \multirow{3}{*}{NO}
& $-\pi/2$ 
& $1.17\times10^{4}$ & $2.25\times10^{4}$ & 52.0
& $3.253\times10^{-1}$ & $4.156\times10^{2}$ & 0.078 \\
&& $0$ 
& $1.16\times10^{4}$ & $2.25\times10^{4}$ & 51.6
& $3.145$ & $4.683\times10^{2}$ & 0.672 \\
&& $+\pi/2$ 
& $1.13\times10^{4}$ & $2.32\times10^{4}$ & 48.7
& $3.581\times10^{-1}$ & $5.576\times10^{2}$ & 0.064 \\

\midrule

\multirow{3}{*}{T2K}
& \multirow{3}{*}{IO}
& $-\pi/2$ 
& $1.11\times10^{4}$ & $2.17\times10^{4}$ & 51.1
& $2.799\times10^{-1}$ & $3.954\times10^{2}$ & 0.071 \\
&& $0$ 
& $1.07\times10^{4}$ & $2.20\times10^{4}$ & 48.6
& $3.286$ & $4.603\times10^{2}$ & 0.714 \\
&& $+\pi/2$ 
& $1.07\times10^{4}$ & $2.26\times10^{4}$ & 47.4
& $3.099\times10^{-1}$ & $5.457\times10^{2}$ & 0.057 \\

\midrule

\multirow{3}{*}{NO$\nu$A}
& \multirow{3}{*}{NO}
& $-\pi/2$ 
& $4.26\times10^{4}$ & $6.12\times10^{4}$ & 69.6
& $1.195$ & $1.118\times10^{3}$ & 0.107 \\
&& $0$ 
& $4.25\times10^{4}$ & $6.08\times10^{4}$ & 69.9
& $1.146\times10^{1}$ & $1.251\times10^{3}$ & 0.916 \\
&& $+\pi/2$ 
& $4.26\times10^{4}$ & $6.13\times10^{4}$ & 69.5
& $1.315$ & $1.486\times10^{3}$ & 0.088 \\

\midrule

\multirow{3}{*}{NO$\nu$A}
& \multirow{3}{*}{IO}
& $-\pi/2$ 
& $3.83\times10^{4}$ & $5.51\times10^{4}$ & 69.5
& $1.010$ & $9.713\times10^{2}$ & 0.104 \\
&& $0$ 
& $3.75\times10^{4}$ & $5.55\times10^{4}$ & 67.6
& $1.209\times10^{1}$ & $1.157\times10^{3}$ & 1.046 \\
&& $+\pi/2$ 
& $3.77\times10^{4}$ & $5.63\times10^{4}$ & 67.0
& $1.132$ & $1.407\times10^{3}$ & 0.080 \\

\bottomrule
\end{tabular}
\end{table}

Table~\ref{tab:merged_fisher} summarizes the total event-level Fisher information, the event-weighted QFI, and the corresponding information extraction efficiency for both experiments in the combined neutrino and antineutrino beam modes. The results are compared for the estimation of both $\theta_{13}$ and $\delta_{\rm CP}$. It is evident that, for both experiments, the event-level and event-weighted QFI for $\theta_{13}$ are significantly larger than those for $\delta_{\rm CP}$ i.e. $F_{\rm Q,event}^{\theta_{13}} \gg F_{\rm Q,event}^{\delta_{\rm CP}}$, across both orderings and for all chosen values of $\delta_{\rm CP}$.  The dependence on the neutrino mass ordering is also mild. Overall, the information extraction efficiency for $\theta_{13}$ is significantly higher than that for $\delta_{\rm CP}$. This indicates that the energy spectrum retains a substantial fraction of the available quantum information for the estimation of $\theta_{13}$ compared to $\delta_{\rm CP}$ even after incorporating detector effects.  

As shown in Table~\ref{tab:merged_fisher}, the event-level Fisher information for $\delta_{\rm CP}$ is around two orders of magnitude smaller than the corresponding event-weighted quantum benchmark, resulting in only percent-level information extraction efficiency across all benchmark scenarios. The reported numerical values are obtained under the assumption that the remaining oscillation parameters are fixed, thereby isolating the information content associated with the parameter being estimated. While these numerical values may vary with the choice of the oscillation parameters and the adopted external priors, the limited information-extraction efficiency observed for $\delta_{\rm CP}$ implies that such variations are not expected to alter the qualitative conclusions of the present work.

\section{Impact of backgrounds and systematics on the estimation of $\delta_{\rm CP}$}
\label{sec:bkgplussys}
In this section, we discuss the impact of backgrounds and systematic  uncertainties on the event-level Fisher information. These effects are  expected to further reduce the accessible information; however, we quantify  the corresponding degradation within our framework. In other words, the  information lost specifically due to these experimental effects can, in  principle, be reduced by improving the experimental design and systematic  control. This approach can therefore identify, or at least provide an  indication of, which backgrounds and systematic uncertainties constitute the dominant sources of information loss.

\subsection{Impact of Backgrounds}
\label{subsecA}
Not all background contributions are independent of the variation in the oscillation parameters. In particular, the WS beam contamination discussed in Sec.~\ref{sec:setup} undergo flavor oscillations and the overall contribution depends on the choice of $\delta_{\rm CP}$. Therefore, we consider WS contribution as a separate component in the expected number of events as
\begin{equation}
N_k(\delta_{\rm CP}) = S_k(\delta_{\rm CP}) + B_k^{\rm WS}(\delta_{\rm CP}) + B_k\;,
\label{eq:sample}
\end{equation}
where $S_k(\delta_{\rm CP})$ denotes the expected signal, $B_k^{\rm WS}(\delta_{\rm CP})$ is the WS beam contamination component and $B_k$ is the remaining background contributions.
\begin{figure}[H]
    \centering
    \includegraphics[width=0.9\linewidth]{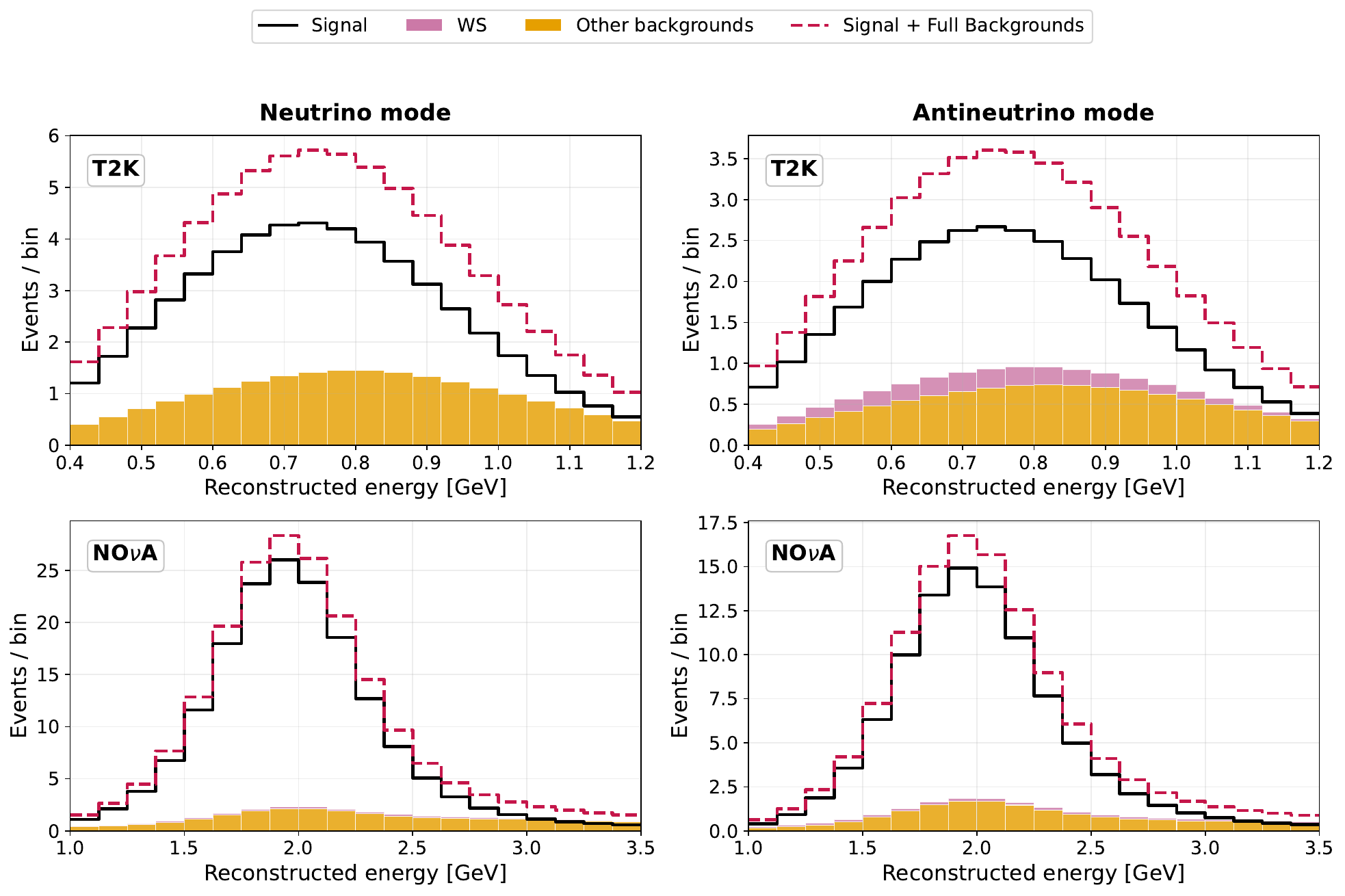}
    \caption{The reconstructed event spectra for T2K and NO$\nu$A in neutrino (left panel) and antineutrino (right panel) modes for NO are shown for $\delta_{\rm CP}=\pi/2$. The corresponding signal (black), oscillated wrong-sign (WS) background (purple), remaining backgrounds (orange), as well as the total signal plus full backgrounds (red) are displayed separately.}
    \label{fig:spectra}
\end{figure}
Figure~\ref{fig:spectra} shows the simulated event spectra for T2K and 
NO$\nu$A. The contributions from the signal (black), WS component (purple), and other backgrounds (orange) are shown separately. It is clearly visible 
from the figure that the WS contribution is more pronounced in T2K, particularly in antineutrino-beam mode, whereas its contribution remains comparatively smaller for NO$\nu$A. Using this spectra, we construct the event-level Fisher information including backgrounds as 
\begin{equation}
F^{\rm tot}_{\rm event} (\delta_{\rm CP}) = \sum_k \frac{1}{N_k(\delta_{\rm CP})} \left[ \frac{\partial S_k (\delta_{\rm CP})}{\partial\delta_{\rm CP}} + \frac{\partial B_k^{\rm WS}(\delta_{\rm CP})}{\partial\delta_{\rm CP}} + \frac{\partial B_k}{\partial\delta_{\rm CP}} \right]^2\;.
\label{eq:event_fisher_background}
\end{equation}

It is worth noting that parameter-independent background contributions enter the event-level Fisher information only through the total event rate in the denominator. In contrast, the parameter-dependent WS contribution, 
$B_k^{\rm WS}(\delta_{\rm CP})$, also enters through its derivative with respect to $\delta_{\rm CP}$ and therefore itself carries information about $\delta_{\rm CP}$. 

The inclusion of WS backgrounds in the event-level Fisher information requires a corresponding modification of the quantum-information benchmark, since the WS events appearing in the reconstructed spectrum originate from a different neutrino quantum state and therefore carry their own quantum  information. Since the expected event sample contains contributions from both neutrino and antineutrino components, the event-weighted QFI should  account for the information carried by each of these states. We therefore treat the signal and the beam component giving rise to the WS background as distinct components and construct the event-weighted QFI by including the  contribution from each corresponding quantum state. Thus, the event-weighted QFI for neutrino and antineutrino state can be written as, 
\begin{equation}
F_{Q}^{\nu(\bar{\nu})}(\delta_{\rm CP}) = \sum_k \left[ C_k^{\nu(\bar{\nu})} F_{Q,\nu(\bar{\nu})}^{\rm pure}(E_k,\delta_{\rm CP}) +
C_k^{\bar{\nu}(\nu),\rm WS} F_{Q,\bar{\nu}(\nu)}^{\rm pure}(E_k,\delta_{\rm CP}) \right]\;, \label{eq:qfi_ws}
\end{equation}
where $C_k^{\nu(\bar{\nu})}$ denotes the unoscillated event spectrum of the signal component, as defined in Eq.~\eqref{eq:unosc}, while $C_k^{\bar{\nu}(\nu),{\rm WS}}$ denotes the corresponding event weight of the unoscillated WS beam component. The latter is obtained by factoring out the $P_{\bar\mu \bar e}$ ($P_{\mu e}$) appearance probability from the expected WS event contribution in neutrino (antineutrino) mode.

\begin{figure}[H]
    \centering
    \includegraphics[width=1\linewidth]{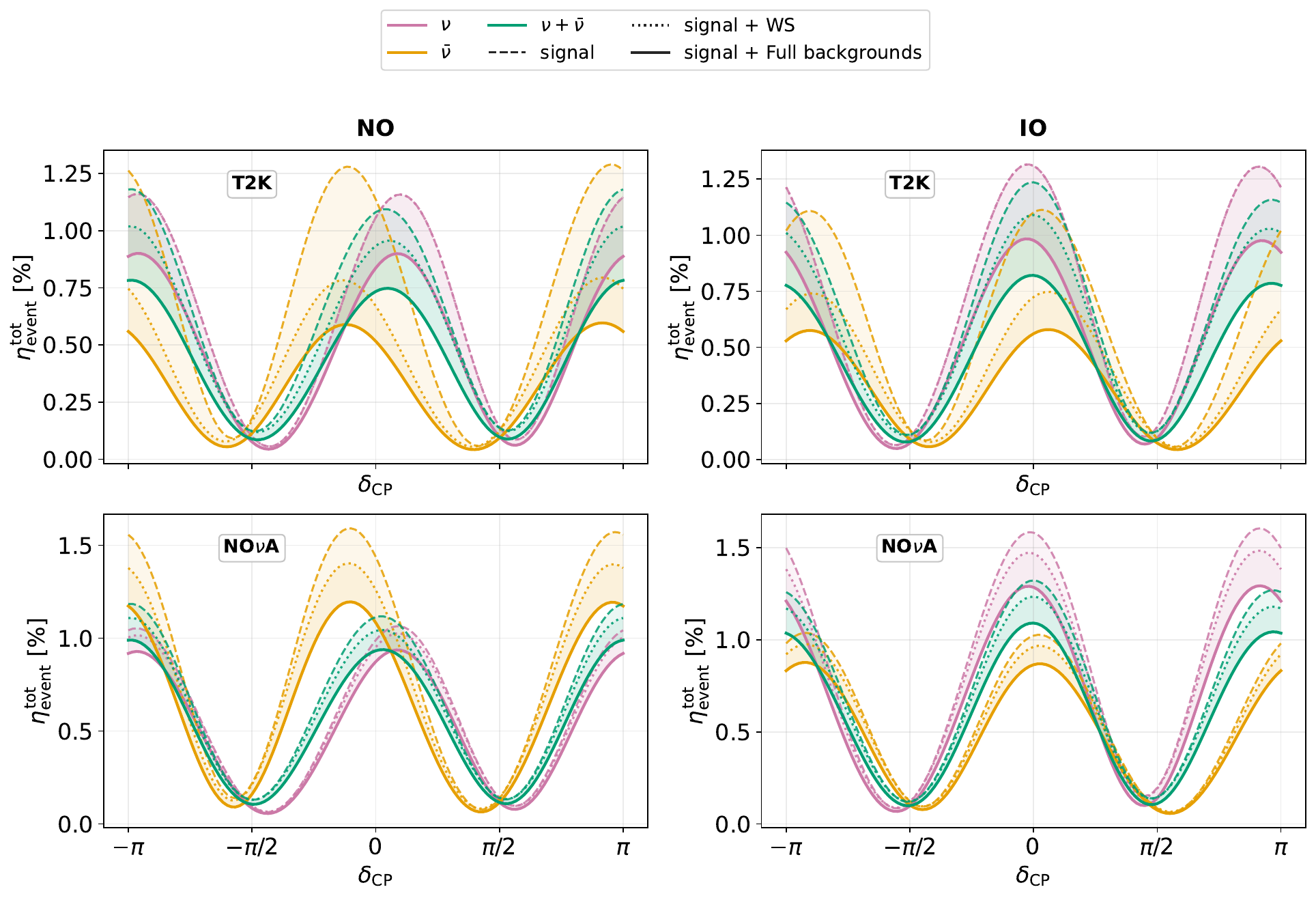}
    \caption{The total information extraction efficiency, $\eta_{\rm event}^{\rm tot}$, for the estimation of $\delta_{\rm CP}$ in T2K and NO$\nu$A considering both NO (left panel) and IO (right panel) is presented. Results are displayed for the neutrino (pink), antineutrino (yellow), and combined $\nu+\bar{\nu}$ (green) modes, successively including the signal only (dashed lines), signal with wrong-sign (WS) component (dotted lines), and signal with full background contributions (solid lines).}
    \label{fig:eta_ws_bg}
\end{figure}

To quantify the impact of information loss due to backgrounds on the  estimation of $\delta_{\rm CP}$, we calculate the total information-extraction  efficiency, $\eta_{\rm event}^{\rm tot}$, by taking the ratio of the  event-level Fisher information defined in Eq.~\eqref{eq:event_fisher_background}  to the corresponding event-weighted QFI given in Eq.~\eqref{eq:qfi_ws}. In Fig.~\ref{fig:eta_ws_bg}, we show the total information-extraction  efficiency as a function of $\delta_{\rm CP}$ after including the background  contributions in the event spectrum. The shaded bands indicate the change in the extraction efficiency resulting from the inclusion of the  corresponding backgrounds, shown separately for the neutrino, antineutrino, and combined beam modes.

It can be seen from the Fig.~\ref{fig:eta_ws_bg} that the WS background component constitutes  a dominant contribution to the information degradation in the antineutrino  mode of T2K for both NO and IO. In all other scenarios, the effect of the  WS background remains non-negligible, with a further degradation observed  after including the remaining background contributions. Thus, the inclusion  of backgrounds in the reconstructed event spectra deteriorates the amount  of information that can be accessed about CP phase.

\subsection{Impact of systematic uncertainties}
\label{subsecB}

We now study the robustness of the event-level Fisher information for the estimation of $\delta_{\rm CP}$ against the systematic uncertainties~\cite{Blennow:2014sja,Huber:2007em,Coloma:2012ji}. In an experiment, the predicted spectrum can be modified due to the uncertainties in the beam normalization and energy scale as they might mimic similar behavior as induced by variations in $\delta_{\rm CP}$.

\subsubsection{Implementation of Normalization uncertainties}
The reconstructed event sample in each energy bin consists of the signal and background contributions as defined in Eq.~\eqref{eq:sample}. To account for normalization uncertainties, we introduce independent nuisance parameters $\xi_s$, $\xi_{\rm WS}$, and $\xi_b$ for the signal, WS background, and other background contributions, respectively. In particular, the normalization of the WS component of backgrounds is allowed to vary independently through $\xi_{\rm WS}$. The expected number of events in the $k$-th energy bin is then given by
\begin{equation}
N_k(\delta_{\rm CP},\boldsymbol{\eta}_{\rm norm})
= (1+\xi_s)S_k(\delta_{\rm CP}) + (1+\xi_{\rm WS})B_k^{\rm WS}(\delta_{\rm CP}) + (1+\xi_b)B_k^{\rm other}\;,
\label{eq:event_norm_sys}
\end{equation}
where $\boldsymbol{\eta}_{\rm norm}=(\xi_s,\xi_{\rm WS},\xi_b)$ denotes the set of normalization nuisance parameters. The normalization uncertainties are chosen as given in Sec.~\ref{sec:setup}. We conservatively assign a normalization uncertainty of $\sigma_{\rm WS}=20\%$ to the WS contribution for both experiments.

The Fisher information matrix in the enlarged parameter space, $\boldsymbol{p}
=(\delta_{\rm CP},\xi_s,\xi_{\rm WS},\xi_b)$, is then written
as~\cite{Leane:2025efj},

\begin{equation}
F_{ij}^{\rm norm}(\boldsymbol{p})
=
\sum_k
\frac{1}{N_k}
\frac{\partial N_k}{\partial p_i}
\frac{\partial N_k}{\partial p_j}
+\left(F_{\rm prior}^{\rm norm}\right)_{ij},
\label{eq:fisher_norm}
\end{equation}
where
\begin{equation}
\boldsymbol{F}_{\rm prior}^{\rm norm}
=
{\rm diag}
\left(
0,\,
\sigma_s^{-2},\,
\sigma_{\rm WS}^{-2},\,
\sigma_b^{-2}
\right).\nonumber
\end{equation}

We use Schur marginalization~\cite{Kitching:2009} to obtain the remaining information on $\delta_{\rm CP}$ after accounting for the normalization uncertainties. The corresponding event-level Fisher information matrix is given by, 

\begin{equation}
F_{\rm event}^{\rm norm}(\delta_{\rm CP})
=
F_{\delta_{\rm CP}\delta_{\rm CP}}
-
\boldsymbol{F}_{\delta_{\rm CP}\eta}
\boldsymbol{F}_{\eta\eta}^{-1}
\boldsymbol{F}_{\eta\delta_{\rm CP}},
\label{eq:fisher_norm_marg}
\end{equation}

where $\boldsymbol{F}_{\eta\eta}$ is the Fisher matrix associated with the additional normalization nuisance parameters.

\subsubsection{Implementation of energy calibration uncertainty}

We now consider the effect of energy calibration uncertainty, which unlike the beam normalization uncertainty that changes the event rate, this can modify the shape of spectra by migrating events in the neighboring bins. We introduce two independent nuisance parameters $\beta_s$ and
$\beta_b$ for the signal and all backgrounds including WS contribution. For $X=S,B$, the shifted spectrum is evaluated as~\cite{Kopp:2006mw},
\begin{align}
X_i(\delta_{\rm CP},\beta_X) &= (1+\beta_X) \left[ \left( X_{j+1}(\delta_{\rm CP})-X_j(\delta_{\rm CP}) \right)(d-j) +
X_j(\delta_{\rm CP}) \right], \label{eq:energy_calibration_C}\\
d &= \beta_X \left( i+\frac{E_{\min}}{\Delta E}+\frac{1}{2} \right)+i, \qquad j=\mathrm{int}(d),
\nonumber
\end{align}
where $\Delta E$ is the constant energy bin width. After the inclusion of calibration uncertainty in Eq.~\eqref{eq:event_norm_sys}, one can write, 
\begin{equation}
\begin{split}
N_k(\delta_{\rm CP},\boldsymbol{\eta}) ={}& (1+\xi_s)S_k(\delta_{\rm CP},\beta_s) + (1+\xi_{\rm WS})B_k^{\rm WS}(\delta_{\rm CP},\beta_b)+ (1+\xi_b)B_k^{\rm other}(\beta_b)\;,
\end{split}
\label{eq:event_all_sys}
\end{equation}
with $\boldsymbol{\eta} =(\xi_s,\xi_{\rm WS},\xi_b,\beta_s,\beta_b)$. We adopt these calibration uncertainties as described in the Sec.~\ref{sec:setup}. The Fisher matrix is evaluated using Eq.~\eqref{eq:fisher_norm} in the extended $\boldsymbol{p}$ including $\boldsymbol{\eta}$. The information on $\delta_{\rm CP}$ after marginalization over all nuisance parameters is obtained in a similar fashion using Eq.~\eqref{eq:fisher_norm_marg} and the resulting Fisher information is denoted by $F_{\rm event}^{\rm norm+cal}(\delta_{\rm CP})$.

\subsubsection{Information loss due to systematics in T2K and NO$\nu$A}
The treatment of experimental effects, such as backgrounds (Sec.~\ref{subsecA}) and systematic uncertainties (Sec.~\ref{subsecB}), within the Fisher-information framework allows us to trace the successive degradation of information. In this way, the framework also provides a quantitative assessment of which experimental effects constitute the dominant limitations on the extraction of information about $\delta_{\rm CP}$ and, consequently, on its determination. The impact of backgrounds on the event-level Fisher information is shown in Fig.~\ref{fig:eta_ws_bg} and discussed therein.

To further quantify the information loss due to systematic uncertainties we define a quantity as, 
\begin{equation}
\mathcal{D}_{\rm norm/norm+cal}(\delta_{\rm CP})[\%] = \left( 1-\frac{F^{\rm norm/norm+cal}_{\rm event}(\delta_{\rm CP})}{F_{\rm event}^{\rm tot}(\delta_{\rm CP})}
\right) \times 100\;,
\end{equation}
where $F^{\rm norm/norm+cal}_{\rm event }(\delta_{\rm CP})$ are the marginalized Fisher information quantities evaluated using Eq.~\eqref{eq:fisher_norm} and $F_{\rm event}^{\rm tot}(\delta_{\rm CP})$ is the event-level Fisher information obtained after the inclusion of backgrounds as given in Eq.~\eqref{eq:event_fisher_background}.

\begin{figure}[H]
    \centering
    \includegraphics[width=1\linewidth]{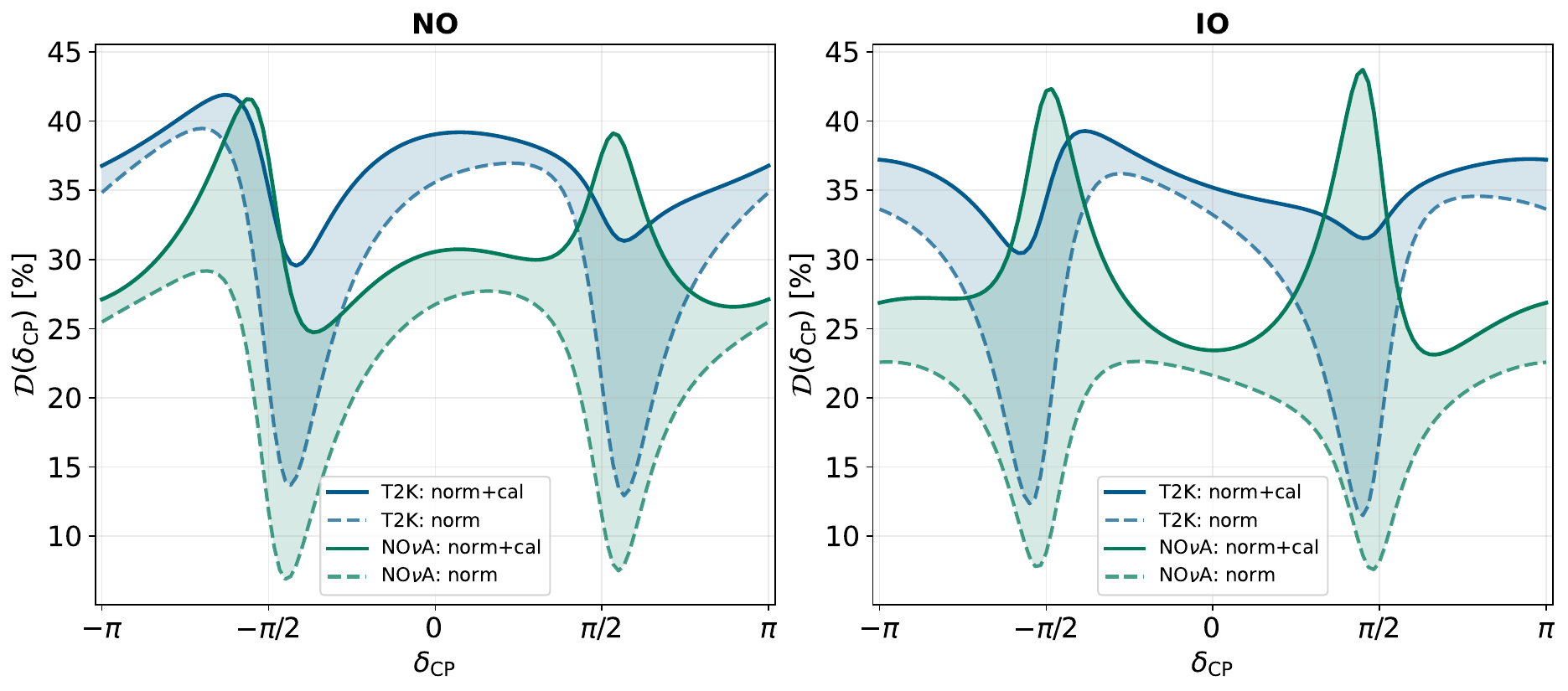}
    \caption{The degradation of the total event-level Fisher information, $\mathcal{D}(\delta_{\rm CP})[\%]$, for the estimation of $\delta_{\rm CP}$ in T2K and NO$\nu$A considering the NO (left panel) and IO (right panel) mass orderings is presented. The dashed curves show the fractional information loss after including and marginalizing over normalization uncertainties, while the solid curves display the corresponding loss after further incorporating a $5\%$ energy-calibration uncertainty.}
    \label{fig:loss}
\end{figure}

We show the fractional loss of event-level Fisher information on $\delta_{\rm CP}$ due to normalization uncertainties, $\mathcal{D}_{\rm norm}(\delta_{\rm CP})$, and after including energy-calibration uncertainties, $\mathcal{D}_{\rm norm+cal}(\delta_{\rm CP})$, in Fig.~\ref{fig:loss}. For both T2K and NO$\nu$A, the information loss exhibits a strong dependence on $\delta_{\rm CP}$ and on the neutrino mass ordering. The normalization uncertainties, shown by the dashed curves, lead to a substantial loss of information near the CP-conserving values of $\delta_{\rm CP}$ compared with the regions around $\delta_{\rm CP}\simeq\pm\pi/2$, reflecting the varying correlations between $\delta_{\rm CP}$ and the normalization nuisance parameters across the parameter space. The solid curves show the information loss after the energy-calibration uncertainties are additionally included, with the separation between the solid and dashed curves representing their further impact beyond normalization uncertainties. This degradation is particularly pronounced in regions where the normalization-induced loss is relatively small, notably around $\delta_{\rm CP}\simeq\pm\pi/2$ for both mass orderings. The complementary impact of normalization and energy-calibration uncertainties therefore leads to a nonuniform degradation of the $\delta_{\rm CP}$ information across its full range, indicating that the effect of systematics cannot be characterized by a single information-loss factor. The information loss due to systematics can also be translated in terms of the total information-extraction efficiency, $\eta_{\rm event}^{\rm tot}$, for instance, at $\delta_{\rm CP}=0$ for NO, the additional percentage loss due to systematic uncertainties is approximately $40\%$
($30\%$) for T2K (NO$\nu$A), compared with about $30\%$ ($15\%$) due to backgrounds, as shown in Fig.~\ref{fig:eta_ws_bg}.

\begin{figure}[H]
    \centering
    \includegraphics[width=1\linewidth]{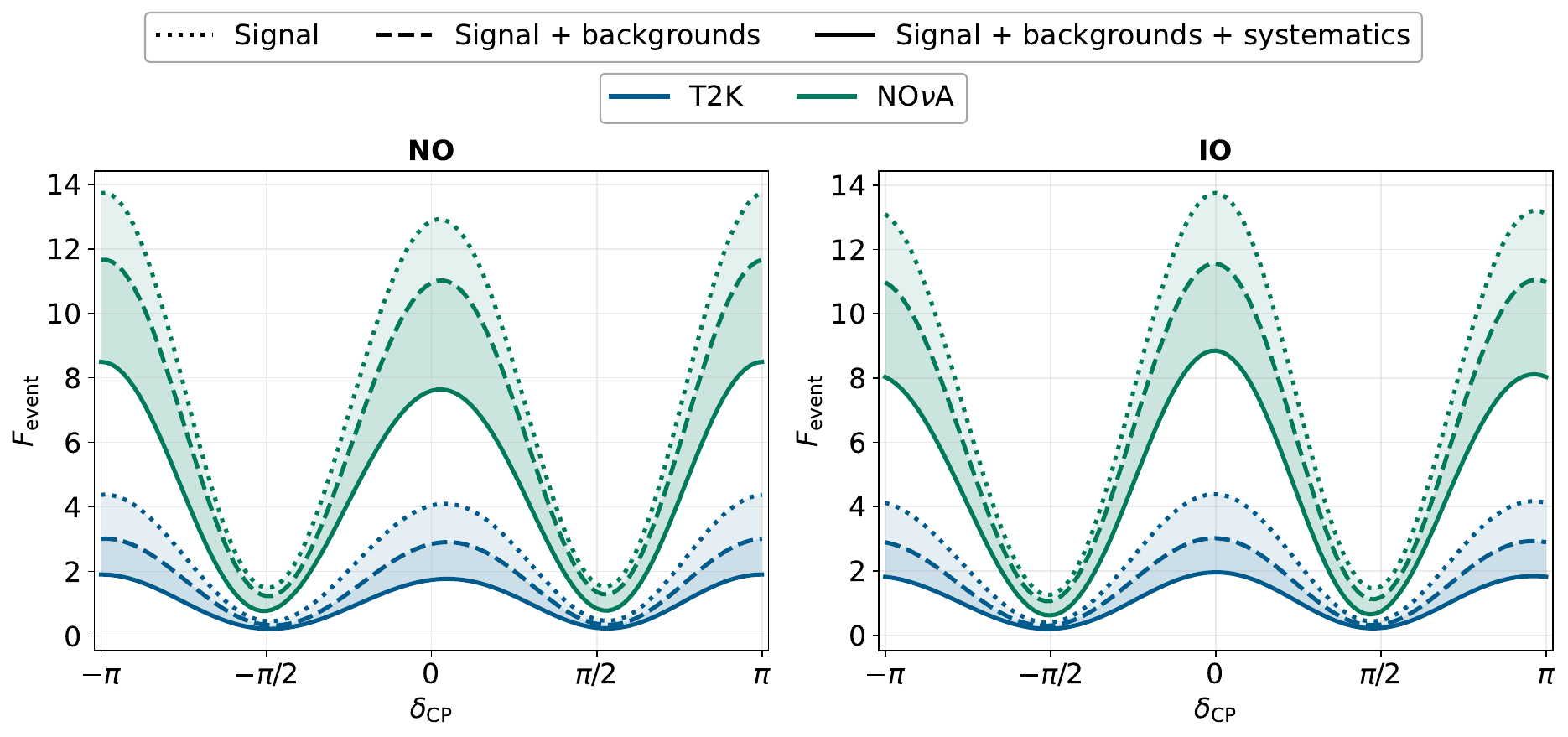}
    \caption{The evolution of the total event-level Fisher information, $F_{\rm event}$, for the estimation of $\delta_{\rm CP}$ in T2K and NO$\nu$A considering the NO (left panel) and IO (right panel) mass orderings is shown. The curves display the successive reduction in $F_{\rm event}$ from the signal-only case (dotted lines) to the inclusion of the full background contributions (dashed lines), followed by the inclusion and marginalization over normalization and energy-calibration systematic uncertainties (solid lines).}
    \label{fig:final-Floss}
\end{figure}

In order to quantify the information losses associated with backgrounds and systematic uncertainties, we show the event-level Fisher information at each successive stage of our analysis in Fig.~\ref{fig:final-Floss}. For T2K, the inclusion of backgrounds in the signal-only case reduces the information on $\delta_{\rm CP}$ by approximately $20$-$30\%$ for both NO and IO. For NO$\nu$A, the corresponding information loss is around $10$-$20\%$ for NO and reaches up to $25\%$ for IO. Thus, T2K exhibits a larger information loss due to backgrounds than NO$\nu$A. A further reduction in information is observed after the inclusion of systematic uncertainties. For T2K, systematic uncertainties reduce the information available for estimating $\delta_{\rm CP}$ by approximately $30$-$40\%$, while for NO$\nu$A the reduction is about $25$-$45\%$, depending on the value of $\delta_{\rm CP}$. The combined effect of backgrounds and systematic uncertainties characterizes the overall event-level Fisher information loss relative to the signal-only case, reaching up to approximately $55\%$ for both experiments and both mass orderings. This also provides an indication of where experimental improvements could minimize the information loss, with backgrounds playing a larger role in T2K, while NO$\nu$A depending on $\delta_{\rm CP}$, being comparatively more affected by systematic uncertainties.

\section{Conclusions and Outlook}
\label{sec:conclusion}
In this work, we study the sensitivity to the leptonic CP-violating phase in the T2K and NO$\nu$A experiments using information theory framework based on Fisher information at both the quantum and statistical event levels, with the primary goal of probing how efficiently the current generation of long-baseline experiments extracts information from measurements compared to the intrinsic quantum information encoded in the propagated neutrino states.

We first investigate the QFI associated with $\delta_{\rm CP}$ separately for neutrino and antineutrino states under both NO and IO. We demonstrate that, for T2K, the difference between the corresponding neutrino and antineutrino states remains comparatively small, whereas for NO$\nu$A the difference becomes significantly larger due to stronger matter effects. This directly indicates that the intrinsic information associated with the oscillation parameter is encoded differently in different neutrino states, and that the interaction with matter during propagation further enhances this distinction. 

For the NO scenario, the neutrino state exhibits comparatively larger QFI than the corresponding antineutrino state, whereas for IO the behavior reverses and the antineutrino state shows larger QFI than the neutrino state. This indicates that the maximum amount of intrinsic quantum information relevant for the estimation of $\delta_{\rm CP}$ is dependent on the assumed neutrino mass ordering for a specific state. However, upon combining the neutrino and antineutrino states, the corresponding QFI becomes nearly identical and largely independent of the choice of mass ordering. We find that the QFI exhibits only a weak dependence on the value of $\delta_{\rm CP}$ for both neutrino and antineutrino states individually, as well as for the combined configuration, under both neutrino mass orderings.

We also define the asymmetry in QFI between the neutrino and antineutrino states, analogous to the conventional probability asymmetry, $A_{\rm CP}$, for the appearance channels of $\nu_e$ and $\bar{\nu}_e$. We show that the QFI asymmetry exhibits a qualitatively similar behavior to $A_{\rm CP}$. For T2K, the corresponding $A_{\rm QFI}$ changes sign and exhibits dependence on the neutrino mass ordering, unlike the probability asymmetry. In contrast, for NO$\nu$A both asymmetries exhibit similar behavior. This further reinforces our observation that the neutrino mass ordering affects not only the oscillation probabilities, but also the intrinsic sensitivity of the propagated quantum states to variations in $\delta_{\rm CP}$.

Next, we include the corresponding event-level spectral information to construct the event-weighted QFI as an experimentally relevant benchmark for the accessible quantum information. We show that the event-weighted QFI develops localized structures around the first oscillation maximum and near the regions $\delta_{\rm CP}=\pm\pi/2$, thereby exhibiting a stronger dependence on $\delta_{\rm CP}$ due to the reconstructed spectral shape in both T2K and NO$\nu$A. The corresponding distributions also exhibit strong beam-mode and hierarchy dependence. For NO$\nu$A, these features become more pronounced due to the sharper spectral peaks around the oscillation maximum and the stronger matter effects during propagation.

Using the standard statistical definition of Fisher information derived from the Poisson likelihood associated with reconstructed event measurements, we obtain the corresponding event-level Fisher information for both the T2K and NO$\nu$A configurations. This event-level Fisher information exhibits strong dependence on neutrino energy, beam polarity, and neutrino mass ordering. We also observe that, for T2K, the information is distributed more broadly across the reconstructed energy bins compared to the corresponding event-weighted QFI distributions. In contrast, this feature is comparatively less pronounced in NO$\nu$A due to its sharper spectral shape around the oscillation maximum. Furthermore, the experimentally accessible Fisher information becomes significantly suppressed near the maximally CP-violating values of $\delta_{\rm CP}$ ($\delta_{\rm CP}\simeq\pm\pi/2$). This indicates that the reconstructed event spectra in both T2K and NO$\nu$A become comparatively less sensitive to small variations of the CP phase in these regions.

To estimate how much information is extracted in the experiments from the intrinsic quantum information present in the propagated neutrino states, we define the quantity namely, bin-wise extraction efficiency, as the ratio between the event-level Fisher information and the corresponding effective quantum benchmark given by the event-weighted QFI. The corresponding distributions follow a behavior qualitatively similar to that of the event-level Fisher information itself. The T2K configuration shows nearly identical behavior for both neutrino mass orderings and beam polarities, whereas NO$\nu$A exhibits a stronger dependence on both the mass ordering and the beam mode. It should also be noted that this quantity nearly saturates the effective quantum benchmark in the very low-energy bins for all four T2K scenarios, while in NO$\nu$A such saturation is observed only for the antineutrino mode in NO and the neutrino mode in IO. We therefore infer that, over most of the $(E_\nu,\delta_{\rm CP})$ parameter space, both experiments are unable to extract the majority of the intrinsic quantum information present in the propagated neutrino states. Around the maximally CP-violating region $\delta_{\rm CP}\simeq -\pi/2$, the corresponding extraction efficiency becomes negligibly small over almost the entire energy range.

We next quantify the total extraction fraction by taking the ratio of the total event-level Fisher information to the corresponding total event-weighted QFI. This quantity serves as an effective detector-level figure of merit for measuring how efficiently the complete reconstructed event spectrum extracts the available quantum information relevant for the estimation of $\delta_{\rm CP}$. We find that neither T2K nor NO$\nu$A fully saturates the corresponding event-weighted quantum benchmark, and both experiments extract only a fraction of the intrinsic quantum information.

For the NO scenario, the maximal total extraction efficiency in the antineutrino mode remains comparatively smaller for T2K than for NO$\nu$A due to the stronger matter effects in the latter. Under IO, the dominant mode for information extraction shifts from the antineutrino channel to the neutrino channel, reflecting the impact of the neutrino mass ordering on the experimentally accessible information. We also show that combining the two beam modes largely removes the distinction between the mass orderings and leads to nearly identical total extraction efficiencies for both T2K and NO$\nu$A.

The present analysis provides a complementary framework for understanding the estimation of the leptonic CP-violating phase in neutrino oscillation experiments. We demonstrate that the sensitivity to $\delta_{\rm CP}$ can be understood not only through conventional statistical approaches, but also through how efficiently the detector extracts information about the parameter from the quantum state. The inclusion of backgrounds and systematic uncertainties further deteriorates this extraction by reducing the event-level Fisher information. To provide a benchmark for comparison, we also perform the same analysis for the precisely measured reactor mixing angle $\theta_{13}$. Both T2K and NO$\nu$A exhibit substantially higher information extraction efficiency for $\theta_{13}$ than for the leptonic CP-violating phase. A more robust understanding can be achieved by extending the present analysis to include all oscillation parameters together with their mutual correlations.

The framework developed in this work provides an approach to quantify the information available to estimate the oscillation parameters in the neutrino oscillation experiments and how they differ from the quantum limit given by the neutrino state. The same methodology can be applied to the present and future neutrino experiments such as JUNO, KM3NeT, Hyper-K, or the IceCube upgrade. However, atmospheric neutrino oscillation experiments, owing to their distinct distributions of baselines and energies and the nature of their initial neutrino ensembles, rather than the controlled beams available in T2K and NO$\nu$A, require a dedicated extension of the present framework. This makes long-baseline accelerator experiments such as T2K and NO$\nu$A a cleaner setting for the present analysis. Such applications will further broaden and confirm the robustness of the proposed framework.

\section*{Acknowledgments} 
NRSC\ and LAD\ were supported in part by the Kaiping Neutrino Research Center, China. YFL\ was supported in part by the National Natural Science Foundation of China under Grant No.~12075255. We also acknowledge the comments and suggestions of the anonymous referee, which helped to improve the clarity of the discussion and presentation of the results in this manuscript.

\end{document}